\documentclass[prb,aps,twocolumn,longbibliography,floatfix]{revtex4-2}
\usepackage[
  colorlinks=true,
  urlcolor=blue,
  linkcolor=blue,
  citecolor=blue,
  bookmarks=false,
  pdftitle={Protected  Weyl semimetals within 2D chiral classes},
  pdfauthor={Faruk}
]{hyperref}

\usepackage{import}
\usepackage{subfiles}
\usepackage{graphicx}
\usepackage{bm}
\usepackage{amsmath}
\usepackage{amssymb}
\usepackage{mathtools}
\usepackage{amsfonts} 
\usepackage[all]{xy}
\usepackage{xspace}
\usepackage{accents}
\usepackage{stmaryrd}
\usepackage{tabularx}
\usepackage{array}
\usepackage{extarrows}
\usepackage{float}
\usepackage{multirow}
\usepackage{makecell}
\usepackage[capitalise]{cleveref}
\usepackage[dvipsnames]{xcolor}
\usepackage[normalem]{ulem}
\usepackage{pdfpages}
\makeatletter
\AtBeginDocument{\let\LS@rot\@undefined}
\makeatother

\newcommand{\comment}[1]{}
\newcommand\beq{\begin{equation}}
\newcommand\eeq{\end{equation}}

\begin{document}

\title{Protected Weyl semimetals within 2D chiral classes}

\author{Faruk Abdulla}
%\email{farukhrim@gmail.com}
%\affiliation{$^1$Harish-Chandra Research Institute, A CI of Homi Bhabha National
%Institute, Chhatnag Road, Jhunsi, Prayagraj (Allahabad)  211019, India}
\affiliation{Physics Department, Technion - Israel Institute of Technology, 
Haifa 32000, Israel}

\begin{abstract}

Weyl semimetals in three dimensions can exist independently of any symmetry apart from
translations. In contrast, in two dimensions, Weyl semimetals require additional 
symmetries, including crystalline symmetries, to exist. Previous research, based on K-theory 
classification, suggested that chiral symmetry can protect Weyl nodes in two dimensions. 
According to K-theory, stable Weyl nodes can exist in four chiral classes—AIII, BDI, CII, 
and DIII—and are classified by $\mathbb{Z}$ (AIII, BDI, DIII) and $\mathbb{Z}_2$ (CII) 
invariants. However, it 
was later found that the $\mathbb{Z}_2$ and trivial indices predicted by K-theory do not reliably 
indicate the presence or absence of Weyl nodes in two dimensions. In this study, we 
demonstrate that stable Weyl nodes exist in each of the five chiral classes and can be 
characterized by a $\mathbb{Z}$ winding number in two dimensions. Our conclusion is supported by the 
explicit solution of the most general Hamiltonian consistent with the symmetry class. We 
also discuss protected Fermi arc edge states, which always connect the projections of Weyl 
nodes with opposite topological charges. Unlike the surface states in three-dimensional Weyl 
semimetals, the edge states in two-dimensional  Weyl semimetals within chiral classes are 
completely dispersionless and remain at zero energy due to the protecting chiral symmetry.

\end{abstract}

\maketitle

%%%%%%%%%%%%%%%%%%%%%%%%%%%%%%%%%%%%%%%%%%%%%%%%

\section{Introduction}

In the recent decades, three dimensional topological Weyl semimetals (WSMs)
\cite{Shuichi_Murakami_2007, Wan_Turner_2011, Burkov_Balents_2011,  Burkov_Hook_2011, 
Xu_Weng_2011, Xu_Belopolski_2015, Xu_Alidoust_2015, Lv_Weng_2015a, Lv_Weng_2015b} have gone 
through an intensive studies owing to their many  exotic properties, such as chiral anomaly 
\cite{Adler_1969, Nielsen_Ninomiya_1983,  Aji_2012, Zyuzin_Burkov_2012},
negative magneto resistance \cite{Son_Spivak_2013, Gorbar_Miransky_2014, Burkov_2015, 
Li_Roy_2016, Lu_Shen_2017, Das_Agarwal_2019, Das_Singh_2020}, quantum oscillation 
and topologically protected Fermi arc 
surface states \cite{Potter_Kimchi_2014, Moll_Nair_2016, Naryan_Watson_2015, Gorbar_Mransky_2016}. 
In 3D WSMs, non degenerate  valence and conduction bands touch at 
a finite number of isolated points in 3D Brillouin zone (BZ), in such a way that the low  energy 
dispersion around each of the touching points is linear in every direction. Thus, when the 
chemical potential  is pinned at the Weyl node (WN),  the low energy  physics in 3D WSMs are governed 
by the relativistic massless Weyl fermions. The  WSM state is the most robust topological gapless 
phase in three dimensions because it does not require any symmetry (except lattice translation) 
for the protection of the  Weyl nodes. 

Considering the novelty of the WSM state and the exotic  phenomena it exhibits in 
three dimensions, it is natural to ask whether WNs can occur in two dimensional 
materials also. Let us first clarify the usage of the terms  WNs and WSMs state in two-dimensional 
systems. Just as in three-dimensional  WSMs, two-dimensional systems 
can exhibit instances where the nondegenerate valence and conduction bands come into 
contact at isolated  points within the two-dimensional  Brillouin zone. These touching 
points within two-dimensional 
systems, akin to the WNs found  in  three-dimensional WSMs, exhibit analogous energy 
dispersion characteristics and possess nontrivial topological charges that safeguard their 
existence. Consequently, these touching  points are  also termed WNs, and materials hosting 
them are referred to as 2D WSMs \cite{Hou_2013, Hou_2014, Young_Kane_2015, Yang_2016, 
Isobe_Nagaosa_2016, Kim_Keun_2017, Sigrist_2017, You_Chen_2019, You_Su_2019, Guo_Lin_2019, He_Yao_2020, Jin_Zheng_2020, Jia_Liu_2020, Feng_Zhu_2021, Shi_Liu_2021, 
Meng_Zhang_2021, Li_Wang_2021, Zou_Niu_2021, Li_Trauzettel_2022, Wei_Tao_2022, 
Lopes_Rogerio_2023, Stemmer_2023, Stemmer_2024, Smith_2024} 
A Dirac semimetal in 2D, where doubly degenerate valence 
and conduction bands touch linearly at isolated points (called Dirac nodes) in 2D BZ, is 
distinguished from a 2D WSM  by the degeneracy of the touching point. 

The presence  of nontrivial topology in the band structure of  semimetals 
and metals can be understood by enclosing  the band touching point(s) in a lower dimensional 
manifold within the Brillouin zone.  By construction  the Hamiltonian, which is 
restricted on the enclosing  manifold is always  gapped,  describes a lower dimensional gapped 
state. Thus, the topological 
categorization of band touching phenomena in metals and semimetals simplifies to the 
classification of topological gapped states found on the lower-dimensional enclosed manifold
\cite{Horava_2005, Ryu_Schnyder_2010, Zhao_Wang_2013, Matsuura_Chang_2013, Zhao_Wang_2014, 
Chiu_Schnyder_2014, Chiu_Ryu_2016}. 
For instance, in the context of 3D Weyl semimetals, a WN is surrounded by a
two-dimensional surface, and the Hamiltonian confined to this surface characterizes 
Chern insulator. Therefore, the WNs in 3D WSMs carry a nonzero Chern numbers which 
protect them from gap opening when subjected to an arbitrary but small perturbation.

Unlike 3D WSM, the reduced dimension require additional symmetries to protect a WN in 
two dimensional materials.  Here we are interested on classification of WNs in 2D 
protected by the nonspatial Altland-Zirnbauer (AZ) symmetries: Time reversal ${\cal T}$, 
particle-hole ${\cal C}$ , and their combined chiral symmetry ${\cal S} = {\cal CT}$, only. 
Previous research, based on K-theory classification 
\cite{Zhao_Wang_2013, Matsuura_Chang_2013, Zhao_Wang_2014}, suggested that
chiral symmetry can protect Weyl nodes in two dimensions. According to K-theory, 
stable Weyl nodes exist in four chiral classes—AIII, BDI, CII, and DIII—and are 
classified by $\mathbb{Z}$ (AIII, BDI, DIII) and $\mathbb{Z}_2$ (CII) invariants. 

However, Chiu and Schnyder \cite{Chiu_Schnyder_2014}, by examining symmetry 
preserving gap opening terms and 
existence of topological invariant, later reported that a $\mathbb{Z}_2$ invariant predicted 
by K-theory cannot protect a Fermi point(or point node) located at off high-symmetry 
points in the BZ, although it can lead to the existence of protected surface states. 
Furthermore, Chiu et al. \cite{Chiu_Ryu_2016}
noted that a trivial index predicted by K-theory cannot guarantee 
existence or nonexistence of a stable Fermi surface (nodal surface for a 
gapless superconductor). 

The authors in Ref. \cite{Chiu_Ryu_2016} observed that if there are stable 
Fermi surface (nodal surface) in a class with a trivial index according to K-theory, 
then it is the $\mathbb{Z}$ invariant of the parent class AIII or A which can protect the Fermi 
surface (nodal surface). 
It is this observation which plays a crucial role in the classification of Fermi 
surface protected by internal symmetry in this work. 
Let us elaborate on this.

Suppose a Hamiltonian in a given symmetry class possesses a Fermi surface (nodal surface) in two 
dimensions. To determine whether this Fermi surface can have topological protection, we need to 
examine the topological classification of insulators or superconductors on the lower-dimensional 
enclosing manifold. For a WN in 2D, the enclosing manifold is a one-dimensional closed loop. For a 
generic enclosing loop surrounding the WN that does not include the ${\bf k}$ and  
$-{\bf k}$ points 
in the BZ, the Hamiltonian restricted to this loop can only possess chiral 
symmetry (no AZ symmetry) if the full Hamiltonian belongs to a chiral class (nonchiral class). 
Therefore, the Hamiltonian restricted to a generic enclosing loop belongs to class AIII (A) for 
chiral classes (nonchiral classes). Since a 1D gapped Hamiltonian in class A has a trivial classification, 
WNs within nonchiral classes are not topologically protected. However, because class A Hamiltonians 
have nontrivial topology in zero dimensions, a stable 1D Fermi surface (nodal line) can exist 
within 2D 
nonchiral classes. On the other hand, WNs within chiral classes can have topological protection 
because a 1D gapped Hamiltonian in class AIII has $\mathbb{Z}$ classification. 

We note that the above argument (as was observed by the authors in 
Ref. \cite{Chiu_Ryu_2016} in connection to find out some of the zero 
in the classification table by K-theory may not necessarily indicate trivial topology) 
does not guarantee existence of WNs in chiral classes in two 
dimensions. It only says that if WNs exist in 2D chiral classes, then they can always be
classified by $\mathbb{Z}$ which is a winding number \cite{Chiu_Ryu_2016}
\begin{align}\label{eq:Winvariant}
W = \frac{1}{2\pi i} \int_C d{\bf k} \cdot \mathrm{Tr}\left(Q^{-1} \nabla_{\bf k} Q \right),
\end{align}
where $C$ is an enclosing loop surrounding a WN in the 2D Brillouin zone.  
Hamiltonian in a chiral class can always be brought to block off-diagonal form 
and the matrix $Q({\bf k})$ in Eq. \ref{eq:Winvariant} is the upper right block. 

Here, we are interested in classification of WNs in two-dimensions only. 
In this work, we are not only providing an alternative method for classification 
of WNs, but also showing explicitly the existence of robust Weyl semimetals 
phase within 2D chiral classes. Unlike previous methods\cite{Zhao_Wang_2013, Matsuura_Chang_2013, 
Zhao_Wang_2014, Chiu_Schnyder_2014, Chiu_Ryu_2016}, our approach \cite{Abdulla_Das_2024} 
is quite straightforward: we write down 
the most general Hamiltonian in the class which we solve to show explicitly that 
all five chiral classes possess a robust Weyl semimetal phase in two-dimension. 
Additionally, our solution also 
confirms that WSM phase is the sole stable gapless phase within 2D chiral classes \cite{Chiu_Ryu_2016}. By `most general'  Hamiltonian, 
we  mean that all coupling terms allowed  within a  given symmetry class are  present 
in the Hamiltonian. The Hamiltonian is also assumed to break all the crystalline 
symmetries. In the earlier paragraphs, we discussed that Weyl nodes 
in 2D chiral classes are characterized by a one-dimensional  $\mathbb{Z}$ invariant, 
specifically a winding number (Eq. \ref{eq:Winvariant}). 
It is important to 
highlight that our approach is quite general and can be readily extended to 
other types of Fermi surfaces (nodal surfaces) in different dimensions. Given 
the fact that insulating states in class A and AIII are classified by a  
$\mathbb{Z}$ invariant, 
we anticipate that Fermi surfaces (nodal surfaces) in any dimension, located 
away from high-symmetry points in the BZ within the AZ class, can  be 
classified solely by  $\mathbb{Z}$ invariants.

By the bulk boundary correspondence,  there exists  topologically protected zero energy edge 
states at the boundary. Like the Fermi arc surface states in 3D WSMs, the zero energy edge  
states (which exist only on that open boundary on which projections of WNs of opposite 
charges do not overlap) in 2D WSMs always connect a pair of WNs which carry 
opposite topological  charges (an example in Fig. \ref{fig:fig2}). 
However there is an important distinction between the surface states of a 3D WSMs and the 
edge states of a 2D WSMs: For WSMs in 3D, the surface states are dispersionless along one 
direction but disperse linearly along the transverse direction, whereas the edge states of 
WSMs in 2D are completely dispersionless due to the reduced dimensionality. 

The fact that the Hamiltonian in classes BDI, CII, CI and DIII  has antiunitary 
time-reversal and 
particle-hole symmetry,  puts a restriction on the minimum number of WNs which can appear 
in the  theory.  Similar to WSMs in 3D,  WNs of 2D WSMs always come in pair so that 
net topological charge remains zero.  The number of WNs in a WSM which belongs to 
BDI and CII  (${\cal T}^2 = {\cal C}^2$) is  always a multiple  of two, whereas in class  CI 
and DIII (${\cal T}^2 =-{\cal C}^2$) the number of WNs must  be  a multiple of four.

Organization of this manuscript  is as follows: In Sec. \ref{Sec:WeylTouching}, 
we briefly review band 
touching in a generic  two bands model to illustrate how WNs can occur in two dimensions
when chiral symmetry is imposed. Then in Sec. \ref{Sec:Class},  we set up  and fully 
explore a minimal model with a four by  four Hamiltonian in every chiral classes and 
show how the generic gapless phase in a given  class  can be obtained. We write 
down lattice models in Sec. \ref{Sec:FermiArc} to demonstrate the Fermi arc edge  
states associated  with the WNs of WSM in two dimensions. We discuss  our findings 
and conclude  in  Sec. \ref{Sec:Summary}. Technical details regarding the transformation 
of the winding  invariant $W$ under time-reversal can be found  in the Appendix \ref{App:A}. 

%%%%%%%%%%%%%%%%%%%%%%%%%%%%%%%%%%%%%%%%%%%%%
\section{Weyl touching in a generic two bands model}
\label{Sec:WeylTouching}

Since Weyl touching, either in 2D or 3D, involves only two non degenerate bands, one can 
consider a generic two bands model,
\begin{align}\label{eq:WeylTwoBand}
H({\bf k}) = f_1({\bf k}) \sigma_x  +  f_2({\bf k}) \sigma_y  +   f_3({\bf k}) \sigma_z,
\end{align}
to see whether a robust Weyl touching is possible between the two bands. We have ignored 
the identity term $f_0({\bf k}) \sigma_0$ because it is not relevant to our analysis below. 
The $\sigma$'s in Eq. \ref{eq:WeylTwoBand} represent Pauli matrices. For touching 
between the two bands, all the three function $f_i({\bf k})$, $i=(1,2,3)$, must vanish simultaneously 
for some ${\bf k} = {\bf k}_0$. This is always possible in 3D because one has three variables
 ${\bf k} = (k_x, k_y, k_z)$ to tune  to make the three function $f_i({\bf k})$ vanish simultaneouesly. If there is 
 no additional constraint on  $f_i({\bf k})$ due to some crystalline symmetry, the touching at ${\bf k}_0$ 
 will be in general linear in every direction. Therefore Weyl points  in 3D  appear without any symmetry 
 requirement except translation. Clearly, any small perturbation $\delta f_i({\bf k}) \sigma_i $ 
 cannot remove the touching point immediately, but can shift the touching point  in the  k-space. 
 Topological nature of 
 the WN at ${\bf k}_0$ is revealed by considering a 2D enclosing surface surrounding the WN in the 
 3D BZ. For a generic enclosing 2D surface, the Hamiltonian restricted on it  belongs to class
 A and a 2D Hamiltonian in this class has nontrivial topology which is  characterized by the Chern number
 \cite{Armitage_Vishwanath_2018} 
 \begin{align}
 C = \frac{1}{2\pi} \int_S d{\bf S} \cdot {\bf F}, 
 \end{align}
 where $S$ is a 2D enclosing surface within the 3D BZ and ${\bf F} = \nabla \times {\bf A}$ is the 
 Berry curvature. So WNs  in three dimensions carry nonzero Chern numbers which provide 
 them topological protection against gap opening. 
 
Coming to the band touching in 2D, we see that it is in general not possible to make 
three function vanish by tuning only two variables ${\bf k} = (k_x, k_y)$. However, if one 
imposes appropriate symmetries on $H({\bf k})$, then one can obtain robust touching 
protected by those symmetries. As it was shown by the earlier works, some form of 
crystalline symmetry (sometimes more than one)  may  stabilize a Weyl point 
in 2D. However if chiral symmetry is imposed i.e.  ${\cal S} H({\bf k}) {\cal S} 
= -H({\bf k})$, then no 
crystalline symmetry is required to protect WNs in two-dimension.  
In presence of chiral symmetry, one of the $f_i({\bf k})$ must vanish identically 
for all ${\bf k} = (k_x, k_y)$. Suppose, we represent the chiral symmetry operation 
by ${\cal S} = \sigma_z$, then the generic two bands  model can have only 
two terms in the  Bloch-Hamiltonian
 \begin{align}
 H({\bf k}) = f_1({\bf k}) \sigma_x  + f_2({\bf k}) \sigma_2. 
 \end{align}
For band touching, the two  function $f_1$, $f_2$ must vanish simultaneously  for some 
${\bf k} = {\bf k}_0$. And this is always possible in 2D because one has two variables
${\bf k} = (k_x, k_y)$ to tune  to make the two function $f_1$, $f_2$ vanish simultaneously.
Clearly, any small perturbation $ \delta h = \delta f_1({\bf k}) \sigma_x + \delta f_2({\bf k})  \sigma_y$ cannot remove 
the touching point immediately but shifts it in the  2D k-space. So in 2D, WNs are generically 
obtained in a chiral symmetric two bands model without requirement of any  crystalline 
symmetry. Topological nature of the WN at ${\bf k}_0$ 
is revealed by considering a 1D enclosing loop surrounding the WN in the 2D BZ. As 
we have discussed in the previous section, WNs in 2D systems (which belongs to 
a chiral class) are characterised by winding number  $W$ (Eq. \ref{eq:Winvariant}). 
We note that, depending on the symmetry class, K-theory \cite{Zhao_Wang_2013, Zhao_Wang_2014, Matsuura_Chang_2013, Chiu_Schnyder_2014, Chiu_Ryu_2016}, finds  both $\mathbb{Z}$ and $\mathbb{Z}_2$ classification for WNs in two-dimensions. 
However, our argument demonstrates that WNs in 2D chiral classes can consistently be classified using a  $\mathbb{Z}$  winding number. This $\mathbb{Z}$ classification also accounts for the boundary states observed in our models.

Note that the chiral symmetry may originate from sublattice, orbital or is simply 
obtained from the combined symmetry of  time-reversal and particle-hole \cite{Ryu_Schnyder_2010}. 
It may also be an emergent symmetry  in a theory \cite{Chen_Young_2015}. 
The above example illustrates the possibility of a 2D WSM with chiral symmetry 
(but no crystalline symmetries) but is too simple to accommodate the full complexity of 
a generic 2D Hamiltonian with spin and orbital degrees of freedom. In what follows, 
we will explicitly show that the generic Hamiltonian belongonh to a chiral class
possesses a robust gapless phase which is a WSM in two-dimension.

%%%%%%%%%%%%%%%%%%%%%%%%%%%%%%%%%%%%%%%%%%%%%
\section{Weyl semimetals within  2D chiral classes}
\label{Sec:Class}

It is known for decades that the transition from a 3D topological gapped phase to a trivial 
gapped phase, within non chiral classes,  passes through an intermediate  gapless phase 
which is a topological WSM in three dimensions. Murakami \cite{Shuichi_Murakami_2007} 
and  Burkov-Hook-Balents \cite{Burkov_Hook_2011}  demonstrated that  a generic
Hamiltonian  in class AII in 3D possesses a stable gapless phase (as an intermediate 
phase between topological  and trivial gapped states) which is generically a 
Weyl semimetal. Similarly, topological nodal line semimetals appear as a stable intermediate 
gapless phase \cite{Sato_Masatoshi_2006, Beri_2010, Abdulla_Das_2024} in the transition from topological to trivial gapped states within the 
3D chiral classes  AIII, DIII, CI and CII.  Interestingly, this nodal line 
semimetal state  which appears as an intermediate gapless phase between topological  
and trivial gapped states  does not require any crystalline symmetry for protection. 

Previous studies assumed  the necessity for the class to have  a nontrivial topological 
gapped phase in three dimensions, in order to have a stable gapless phase as 
an intermediate  state in the transition from topological to trivial gapped states. 
However, this is not  a necessary condition. A general Hamiltonian within a given 
class in a given dimension, which has trivial topology in that dimension,  can 
still possesses a stable gapless phase with nontrivial  topology.  Here we focus on 
the chiral classes in two dimensions.

We want to elucidate our definition of a  stable or robust phase. Whenever we use 
the term ``stable/robust phase",  we are referring to a phase that 
exists in a robust region of the parameter space. 
It is straightforward to argue that if the generic Hamiltonian within the chiral classes 
possesses  a stable gapless phase, then it  cannot be a Dirac semimetal. Since our 
Hamiltonian has all the terms in the symmetry class, the inversion and the other crystalline 
symmetry are in general broken. Therefore the gapless phase cannot be a Dirac semimetal
which necessarily  requires crystalline symmetry for protection \cite{Young_Kane_2015}. It 
cannot be 
a protected nodal loop semimetal either: Suppose it is a topologically protected nodal loop
semimetal. Then the Hamiltonian restricted on an enclosing manifold must be topologically 
nontrivial.  For a nodal line in 2D BZ, the enclosing surface is  a zero dimensional  manifold 
which consists of only two  points: one inside and the other one outside the nodal loop.  In 
general the Hamiltonian, which is restricted on this manifold,  has chiral symmetry only. 
Therefore,  the Hamiltonian restricted on the enclosing manifold  belongs  to class AIII. 
Since 0D Hamiltonian in  class AIII has trivial topology, the gapless phase  cannot  be a  
stable nodal loop semimetal. Therefore we assert  that the gapless phase (if exists) must be 
a  WSM in two dimensions and as we have discussed in previous sections, the WNs within the 2D chiral classes  are classified by the $\mathbb{Z}$ winding numbers $W$ 
(Eq. \ref{eq:Winvariant}).  

% Reason for the WPs in  any of the five chiral classes can be classified by a single 
% topological invariant $W$ may be understood through  the following argument:  Suppose 
% there is a WP in the theory. 
% Its topology can be revealed by investigating the lower dimensional  insulator (1D insulator)
% which is  living on an enclosing  manifold surrounding the  Weyl point. For a WP in a 2D BZ, the 
% enclosing manifold is a closed loop surrounding the Weyl point. For a generic enclosing loop,
% the ${\bf k}$ and $-{\bf k}$ points of the 2D BZ are not included. This implies that 
% the Hamiltonian which is restricted on a generic enclosing loop will not respect 
% the anti-unitary time-reversal and the particle-hole symmetry of the full Hamiltonian
% in the class. So the Hamiltonian which is restricted on a generic enclosing loop belongs 
% to class AIII. We know that AIII has nontrivial topology in one dimension: The 1D insulators 
% in class AIII are classified by a $\mathbb{Z}$ topological invariant which is the winding number $W$ defined in  Eq. \ref{eq:Winvariant}. Therefore, the WPs (if exist) in the chiral  classes can  
% always be classified by a universal  topological  invariant $W$. 

Our calculation scheme is very similar to those  presented in the Ref. 
\cite{Abdulla_Das_2024}. The goal is to initially write down the most general, 
spinful, $4 \times 4$  
Hamiltonian  within each chiral classes.  Subsequently, we proceed to calculate the
energy spectrum, with a particular focus on solving for zero energy, corresponding to the 
occurrence of band touching. In all the chiral classes, we explicitly show that the condition 
for zero energy  always consists of two independent equations which involve momenta and 
the parameters  in the Hamiltonian. Then we apply our  arguments given in the previous 
section that we have  two momenta $k_x$, $k_y$ in 2D, which we can tune to satisfy two 
equations whose solution space  generically describe a WSM in two-dimension. After 
establishing the existence of 2D WSM phase, we verify  topological protection of 
the WNs  by computing their topological charges  $W$ using Eq. \ref{eq:Winvariant}.

\subsection{Class AIII}

To write down a generic spinful Hamiltonian in class AIII,  let us begin with a simple
(massive) Dirac Hamiltonian 
\begin{align}\label{eq:H0AIII}
    H_0({\bf k}) = \eta_x \otimes \left( k_x \sigma_x + k_y \sigma_y \right) + m \eta_y.
\end{align}
Here $\sigma$'s and $\eta$'s, are two by two Pauli matrices, act on spin and pseudo spin 
degrees of freedom, respectively.
In addition to the chirality  ${\cal S}=\eta_z$,  this Hamiltonian is also symmetric under the 
time-reversal ${\cal T}=i \eta_x \sigma_y  {\cal K}$ (${\cal K}$ complex conjugates anything 
to its right). We must break this time-reversal symmetry to obtain Hamiltonian in class AIII. 
The Hamiltonian in Eq.  \ref{eq:H0AIII} has other crystalline symmetries as well. For instance, 
it has inversion symmetry realized by  ${\cal I}=\eta_y$, and mirror symmetries in the axes 
directions realized  by ${\cal M}_i=\eta_y \sigma_i$, $i=x, y$. Below, in the process of constructing 
the generic Hamiltonian in class AIII, we will break  the crystalline symmetries.

The Hamiltonian in class AIII preserves only chiral symmetry.  There are five more terms which 
can be added to $H_0({\bf k})$ without breaking the chiral symmetry:  $\eta_x$, $\eta_x \sigma_z$
and $\eta_y  \sigma_i$ ($i=x, y, z$). Therefore the most general 
Hamiltonian in class AIII is 
\begin{align}\label{eq:HAIII}
H_{\textrm{AIII}}({\bf k}) = H_0({\bf k}) + \tilde{m} \eta_x \sigma_z + V\eta_x + 
                                          \eta_y \otimes {\bf A} \cdot {\boldsymbol \sigma},                             
\end{align}
where $\tilde{m}$, $V$ are real scalar and ${\bf A} = (A_x, A_y, A_z)$ is a real vector so that 
the Hamiltonian remain hermitian. Note that the terms  $\tilde{m}\eta_x \sigma_z$ and 
$\eta_y \otimes {\bf A} \cdot {\boldsymbol \sigma} $ make sure that time reversal is broken and 
the Hamiltonian in Eq. \ref{eq:HAIII} belongs to class AIII. Further the terms $\tilde{m}\eta_x \sigma_z$ 
and $V\eta_x$ break inversion as well as mirror symmetries. Since crystalline symmetries are 
irrelevant in our discussion, the three scalars $m, \tilde{m}$, $V$ and the vector  ${\bf A}$ can 
be any function of momenta $k_x, k_y$.

To show that the Hamiltonian $H_{\textrm{AIII}}$ in Eq. \ref{eq:HAIII} possesses a 
robust gapless phase, first we have to find its spectrum and then solve it for band 
touching in the six dimensional parameter space ${\bf X} = (m, \tilde{m}, V, {\bf A})$. 
Since the Hamiltonian has chiral symmetry,  the band touching  can occur  at the zero 
energy only. To find the spectrum 
of $H_{\textrm{AIII}}$, we square it, collect the terms and repeat this until the resulting Hamiltonian 
becomes proportional to identity. We find the following energy spectrum,
\begin{align}
E^2({\bf k}) = \tilde{\bf k}^2 + {\bf A}^2 + V^2 + m^2  \pm 2 \sqrt{(V\tilde{\bf k} + m {\bf A})^2 + 
(\tilde{\bf k} \times {\bf A})^2 }, 
\end{align}
where $\tilde{\bf k} = (k_x, k_y, \tilde{m})$ and $\tilde{\bf k}^2 = k_x^2 + k_y^2 + \tilde{m}^2$. 
For band touching (zero energy solution), we must have 
\begin{align}
\left(\tilde{\bf k}^2 + {\bf A}^2 + V^2 + m^2\right)^2  = 4 \left( (V\tilde{\bf k} + m {\bf A})^2 + 
(\tilde{\bf k} \times {\bf A})^2 \right), 
\end{align}
which can be rewritten as 
\begin{align}
\left(\tilde{\bf k}^2 + m^2 - {\bf A}^2 - V^2 \right)^2 + 4 \left(\tilde{\bf k} \cdot {\bf A} - mV \right)^2 = 0. 
\end{align}
Clearly  the zero energy solutions, which are given by the the intersection of two curves 
$k_x^2 + k_y^2 + \tilde{m}^2 + m^2  = {\bf A}^2 + V^2$ and  $A_1 k_x + A_2 k_y =  mV - \tilde{m} A_3$
in the $k_x$-$k_y$ space,  generically describe point touchings (two fold degenerate).
These touching points are the WNs in the theory. To see the zero energy solutions 
describe a stable WSM phase, suppose there is a WN at ${\bf k} = {\bf k}_0$ for a given
choice of the parameters ${\bf X} = {\bf X}_0$. Given the smoothness of the curves in 
parameters (only polynomial dependence), the solution for WNs should  exist for any 
arbitrary but small  change in parameters  $\delta{\bf X}_0$. A  small change in parameters 
${\bf X}_0 + \delta{\bf X}_0$ 
cannot gap out a WN immediately but only can shift  its location to ${\bf k}_0 + 
{\bf \delta k}_0$. 
Therefore we have established  that the  generic Hamiltonian in class AIII  
possesses a robust gapless semimetal phase which is a WSM in two dimensions.

As previously discussed, the WSM  phase we have identified, has  topological protection
which is revealed by computing the topological charge $W$ of the WNs  using 
Eq. \ref{eq:Winvariant}. When the parameters ${\bf X}$ are taken momentum 
independent, there exists only two Weyl points. Numerical computations of the invariant 
$W$ reveal that these two WNs bear opposite topological charges  $W=\pm 1$.  Given their 
nontrivial topological charge, perturbations within the symmetry class cannot eliminate 
them unless two WNs with opposing topological charges approach each other and 
undergo pairwise annihilation.

\subsection{Class BDI}
A generic spinful Hamiltonian in 2D belonging to class BDI  has time-reversal ${\cal T}$, 
particle-hole ${\cal C}$ and  their product ${\cal S}={\cal T} {\cal C}$ preserved with 
${\cal T}^2 = {\cal C}^2 = 1$.  To write down a generic Hamiltonian in  this class, let us 
begin with a massive Dirac Hamiltonian (as we did for the class AIII )
\begin{align}\label{eq:H0BDI}
    H_0({\bf k}) = \eta_y \otimes \left( k_x \sigma_x + k_y \sigma_z \right) + m \eta_x.
\end{align}
Here $\sigma$'s and $\eta$'s, are two by two Pauli matrices, act on spin and pseudo spin 
degrees of freedom respectively. The time-reversal and the particle-hole
operators are realized by  ${\cal T} = \cal{K}$ and ${\cal C} = i \eta_z {\cal K}$ respectively, 
obeying ${\cal T}^2 =1$,  ${\cal C}^2=1$.  Note the Hamiltonian in Eq. \ref{eq:H0BDI} is not the 
most general Hamiltonian in class BDI. There are three additional terms which can be 
added without altering time-reversal and particle-hole symmetry: a scalar $\delta = \eta_y \sigma_y$, 
and a vector ${\boldsymbol \alpha} = (\eta_x \sigma_x, \eta_x \sigma_z)$. Now we can write down 
the most general four by four Hamiltonian in class BDI, 
\begin{align}\label{eq:HBDI}
H_{ \textrm{BDI }}({\bf k}) =  \eta_y \otimes \left( k_x \sigma_x + k_y \sigma_z \right) + m \eta_x + 
 \tilde{m} \delta + {\bf A} \cdot {\boldsymbol \alpha}, 
\end{align}
where ${\bf A} = (A_1, A_2)$ is a two components real vector and $\tilde{m}$ is a real scalar. 
The set of parameters ${\bf X} = (m, \tilde{m}, {\bf A})$ can only be an even function of momenta 
to keep the time-reversal and particle-hole symmetry preserved. 
Without loss of generality, we can consider the following quadratic dependence on momenta: 
$ X_{\alpha}({\bf k}) = X_{\alpha 0} - \sum_{ij} t^{(X_\alpha)}_{ij} k_i k_j$, where $X_{\alpha}$, 
$\alpha=1,2, ..4$, is  a component of the four-dimensional vector ${\bf X}$, and $X_{\alpha0}$, 
$t^{(X_\alpha)}_{ij}$  are all assumed real so that the Hamiltonian remains hermitian. 
It is easily checked that the crystalline symmetries, including inversion and mirrors, 
are broken. 

Now by computing the energy spectrum and solving it for zero energy, we can explicitly 
show whether  the Hamiltonian $H_{\textrm{BDI}}({\bf k})$ in Eq. \ref{eq:HBDI} possesses 
a stable gapless  phase. Repeating the same calculation as we did for class AIII Hamiltonian,
we find  the following energy spectrum of $H_{\textrm{BDI}}({\bf k})$, 
\begin{align}\label{eq:EBDI}
E^2({\bf k}) = {\bf k}^2 + \mu^2 + {\bf A}^2 \pm 2\sqrt{\mu^2 {\bf A}^2 + ({\bf k} \times {\bf A})^2},
\end{align}
where $\mu^2 = m^2 + \tilde{m}^2$. The condition for band touching (or zero energy) is easily obtained: 
\begin{align}\label{eq:E0BDI}
\left({\bf k}^2 + \mu^2 - {\bf A}^2 \right)^2 + 4 \left({\bf k} \cdot {\bf A} \right)^2 = 0. 
\end{align}
Clearly  the zero energy solutions, which are given by the the intersection of two smooth curves 
(i) ${\bf k}^2 + \mu^2 = {\bf A}^2$ and (ii) ${\bf k} \cdot {\bf A}  = 0$ in the k-space, generically 
describe point touchings i.e. Weyl points. Further from the smoothness of the two 
curves in parameters, it follows that  for a given choice of the parameters ${\bf X}  = {\bf X}_0$, 
if ${\bf k} = {\bf k}_0$ is a solution, then a small change in the parameters ${\bf X}_0 + \delta {\bf X}_0$ 
cannot remove the touching point immediately but can shift it in the k-space to ${\bf k}_0 + \delta {\bf k}_0$. 
This implies that there exists a finite region for solution of Weyl points  in the 
parameter space. Therefore  (like in class AIII)  the generic  Hamiltonian in class BDI 
also possesses a stable  gapless  phase which is a WSM in two-dimension.  

The WSM phase has topological protection which can be verified by computing the 
topological charge  $W$ of the WNs by surrounding them in a closed loop. When the parameters 
${\bf X}$ are considered independent  of momenta, the location of the WNs can be easily obtained 
by determining the intersection  of the circle ${\bf k}^2 + \mu^2 = {\bf A}^2$ with the 
straight line  $A_1 k_x + A_2 k_y = 0$. Clearly there are only two WNs which are time-reversal partner of each other. We have  computed the topological charge $W$ of this pair, and  found they
carry opposite topological charges $W=\pm 1$.  Non zero topological charges make the WNs 
topologically stable. The only way to gap out a WN is to approach  another WN of  opposite 
topological charge,  causing them to mutually annihilate. 

\subsection{Class CII}
Hamiltonian belonging to  the symmetry class CII  possesses time-reversal ${\cal T}$, 
particle-hole ${\cal C}$  and the chiral ${\cal S}$ symmetry with ${\cal T}^2=-1, 
{\cal C}^2=-1$. Following the similar steps as we did for class AIII and BDI, the 
most general four by four, spinful, Hamiltonian  in class CII is 
\begin{align}\label{eq:HCII}
H_{\textrm{CII}}({\bf k}) =  \eta_x \otimes \left( k_x \sigma_x + k_y \sigma_z \right) + V_1 \eta_x 
                                           + V_2 \eta_y \sigma_y +   {\bf A} \cdot {\boldsymbol \alpha}, 
\end{align}
where ${\boldsymbol \alpha} = (\eta_y \sigma_x, \eta_y \sigma_z)$,  ${\bf A} = (A_1, A_2)$  is
a two component real vector, and $V_1$, $V_2$ are real scalars. 
Like in class BDI, the set of parameters ${\bf X} = (V_1, V_2, {\bf A})$ can be any 
even function of momentum without breaking the time-reversal and particle-hole symmetry and we can choose them to be quadratic in momenta: $ X_{\alpha}({\bf k}) = X_{\alpha 0} - \sum_{ij} t^{(X_\alpha)}_{ij} k_i k_j$, where $X_{\alpha}$,  $\alpha=1,2, ..4$, is  a component 
of the four-dimensional vector ${\bf X}$, and $X_{\alpha0}$,  $t^{(X_\alpha)}_{ij}$  
are all assumed real. For a generic choice of $t^{(X_\alpha)}_{ij}$, 
it is  easily checked that  crystalline symmetries,  including inversion and mirrors, are broken. 
The energy spectrum is given by
\begin{align}\label{eq:ECII}
E^2({\bf k}) = {\bf k}^2 +  {\bf A}^2  + V^2  \pm 2 \sqrt{V^2 {\bf k}^2 + ({\bf k} \times {\bf A})^2}, 
\end{align}
where $V^2 = V_1^2 + V_2^2$. We note that this energy spectrum is identical to the 
energy spectrum of $H_{\textrm{BDI}}({\bf k})$ given in Eq. \ref{eq:EBDI} with the 
replacement  $\mu^2 \to V^2$. So, the arguments  given after Eq. \ref{eq:E0BDI} for 
the BDI class,  apply to this case as well. Therefore, we arrive at the conclusion that 
a generic Hamiltonian in class CII possesses a robust gapless phase which is WSM 
in two dimensions.

We have confirmed  topological robustness of the WNs by calculating their 
topological charges $W$. Our results indicate that the WNs carry non-zero topological 
charges. Notably, pairs of WNs serving as time-reversal partners exhibit opposite 
topological charges $W= \pm 1$. 

%%%%%%%%%%%%%%%%%%%%%%%%%%%%%%%%%%%%%%%%%%%%%%%%

\subsection{Class CI}
Hamiltonians belonging to class CI must have time-reversal ${\cal T}$, 
particle-hole ${\cal C}$, and the chiral ${\cal S}$ symmetry with  
${\cal T}^2=1, {\cal C}^2=-1$. To write down a generic Hamiltonian in  
this class, let us follow our routine procedure to  
begin with a four by four Dirac Hamiltonian 
\begin{align}\label{eq:H0CI}
    H_0({\bf k}) =  \left( k_x \eta_x + k_y \eta_y \right) \otimes \sigma_y. 
\end{align}
Here $\sigma$'s and $\eta$'s, are two by two Pauli matrices, act on spin and pseudo spin 
degrees of freedom, respectively as it was for the other classes. The time-reversal and the 
particle-hole operators are realized by  ${\cal T} = \eta_x \cal{K}$ and ${\cal C} =-i \eta_y {\cal K}$ 
respectively,  obeying ${\cal T}^2 =1$,  ${\cal C}^2=-1$. The chirality operator is obtained from 
the product of time-reversal and particle-hole: ${\cal S} = {\cal T C} = \eta_z$. Note that 
there is no mass term (which anticommutes with the kinetic terms) for the $4 \times 4$ 
Dirac Hamiltonian in class CI. This was also the case for class CII. The four fold degenerate 
Dirac point located  at $k_x=k_y=0$  is a fine tuned point. We will shortly see when the remaining 
terms allowed by the symmetry class are added, the  Dirac point immediately goes to  Weyl points. 

To write down the most general Hamiltonian in the class, we have to find all the terms allowed 
by the symmetry  class and add them to  $H_0({\bf k})$ in Eq. \ref{eq:H0CI}. We find there are 
six such terms which can be arranged as  two scalars and two  vectors (two components): $\eta_x$, $\eta_y$, 
${\boldsymbol \alpha} = (\eta_x \sigma_x, \eta_x \sigma_z)$ and   ${\boldsymbol \beta} = (\eta_y 
\sigma_x, \eta_y \sigma_z)$. The most general four by four Hamiltonian in class CI is then 
\begin{align}\label{eq:HCI}
H_{\textrm{CI}}({\bf k}) = & \left( k_x \eta_x + k_y \eta_y \right) \otimes \sigma_y + V_1 \eta_x + V_2 \eta_y \\ \nonumber 
& + {\bf A} \cdot {\boldsymbol \alpha} + {\bf B} \cdot {\boldsymbol \beta}, 
\end{align}
where $V_1$, $V_2$ are scalars and ${\bf A}=(A_1, A_2)$, ${\bf B}=(B_1, B_2)$ are 
two components vectors. It is easily checked that the Hamiltonian $H_{\textrm{CI}}({\bf k})$ respects both time-reversal ${\cal T} H_{\textrm{CI}}({\bf k}) {\cal T}^{-1} = H_{\textrm{CI}}(-{\bf k})$ and particle-hole ${\cal C} H_{\textrm{CI}}({\bf k}) {\cal C}^{-1} = 
-H_{\textrm{CI}}(-{\bf k})$ symmetry implemented by  ${\cal T} = \eta_x \cal{K}$  
and ${\cal C} =-i \eta_y {\cal K}$  respectively,  obeying ${\cal T}^2 =1$,  
${\cal C}^2=-1$).  The set of parameters ${\bf X} = \left(V_1, V_2, {\bf A},  {\bf B} 
\right)$ can be any even 
function of momenta without altering the time-reversal and the particle-hole symmetry. 
Without loss of generality,  we can choose the following quadratic dependence on  
momenta: $ X_{\alpha}({\bf k}) = X_{\alpha 0} - \sum_{ij} t^{(X_\alpha)}_{ij} k_i k_j$, 
where $X_{\alpha}$,  $\alpha=1,2, ..6$, is  a component of the six-dimensional vector ${\bf X}$, and  $X_{\alpha0}$,  $t^{(X_\alpha)}_{ij}$  are all real so that the Hamiltonian remains
hermitian.
 
Following the routine steps as before, we get the following  condition for zero energy 
(band touching) 
\begin{align}
\left(\tilde{k}^2 - \tilde{V}^2 + {\bf A}^2 - {\bf B}^2 \right)^2 + \left(k_x k_y + 
{\bf A} \cdot {\bf B} - V_1 V_2 \right)^2 = 0
\end{align}
where $\tilde{k}^2= k_x^2 - k_y^2$ and $\tilde{V}^2 = V_1^2 -V_2^2$. Since the sum of two 
squares is zero, they must vanish separately 
\begin{subequations}
\begin{align}
 k_x^2 - k_y^2  & = {\bf B}^2 + \tilde{V}^2  - {\bf A}^2   \\
 k_x k_y  & = V_1 V_2 - {\bf A} \cdot {\bf B} . 
\end{align} 
\end{subequations}
Similar to the preceding instances within class AIII, BDI, and CII, we observe that the 
solution space, which is given  by the intersection of two smooth curves in the 2D k-space,  
generically describes Weyl touchings and the solution for Weyl points 
exists a finite region in the parameter space ${\bf X}$.   
When the set of parameters ${\bf X}$ are taken independent of momenta, the locations of WNs
are given by the intersection of two hyperbolas. Note that the hyperbola represented by 
the first condition is symmetric about both the $k_x$ and $k_y$ axes, where the second hyperbola
is symmetric about the straight line $k_y = \pm k_x$. Intersections can occur only at two points (two WNs). The WNs are time-reversal partner of each other. Computing  their topological 
charges $W$, we find that they carry identical charges $W=-1$.  Since the WNs carry 
nontrivial topological charges, they cannot be gapped out by  small perturbations unless 
two WNs of opposite charges come close and annihilate each other pairwise.

The CI model in Eq. \ref{eq:HCI} features a pair of WNs with a total charge of -2. 
However, as discussed in later Sec. \ref{Sec:FermiArc}, in a lattice system, the total 
charge must be zero. In a 2D lattice model of a WSM belonging to class CI, a 
minimum of four Weyl nodes is required, with one pair carrying a charge of -W 
and the other +W, resulting in an overall zero charge.

%%%%%%%%%%%%%%%%%%%%%%%%%%%%%%%%%%%%%%%%%%%%%%%%%%%%%%%%%%%%%%%%%%%%%%%%%%%%%%%%%%%%%%%

\subsection{Class DIII}
Hamiltonian that belongs to AZ symmetry class DIII must have time-reversal (${\cal T}$), 
particle-hole (${\cal C}$) and the chiral (${\cal S}$) symmetry and they should obey 
${\cal T}^2=-1, {\cal C}^2=1$. The most general four by four, spinful, Hamiltonian 
in class DIII is
\begin{align}\label{eq:HDIII}
H_{\textrm{DIII}}({\bf k}) =  \eta_x \otimes \left( k_x \sigma_x + k_y \sigma_y \right) + m \eta_y  
                                           + V \eta_x, 
\end{align}
where $m$, $V$ are two real scalars. The time-reversal and the 
particle-hole operators are implemented  by  ${\cal T} = -i \eta_x 
\sigma_y \cal{K}$ and ${\cal C} =  \eta_y \sigma_y {\cal K}$ 
respectively,  obeying ${\cal T}^2 =-1$,  ${\cal C}^2=1$. The chirality 
operator is ${\cal S} = {\cal T C} = \eta_z$. The 
inversion and mirrors symmetries are broken due to the term $V\eta_x$. Any left over crystalline 
symmetry can be removed by considering $V$ and $m$  an arbitrary but even function of 
momenta. The evenness of $V({\bf k})$ and $m({\bf k})$  is required to keep  the 
antiunitary time-reversal and particle-hole symmetry preserved. Without loss of generality, 
we can consider the following quadratic dependence on 
momenta:  $ m({\bf k}) = m_0 - \sum_{ij} t^{(m)}_{ij} k_i k_j$, $V({\bf k}) = V_0 - \sum_{ij}
 t^{(V)}_{ij} k_i k_j$,  where $m_0$, $V_0$,  $t^{()}_{ij}$ are all real. 

The energy spectrum is  found to be, 
\begin{align}
E^2({\bf k}) = {\bf k}^2 + m^2({\bf k}) + V^2({\bf k}) \pm 2 \sqrt{V^2({\bf k}) m^2({\bf k})}. 
\end{align}
The condition for band touching (or zero energy) is readily obtained:
\begin{align}
k_x^2 + k_y^2 = V^2({\bf k}), ~~~~  \sum_{ij} t^{(m)}_{ij} k_i k_j = m_0. 
\end{align}
Clearly, the solution space, which is given by the intersection of two
smooth curves in the 2D k-space,  generically describes Weyl touchings. From the 
smoothness of the  curves, it  follows that the solution for Weyl nodes must exist in a 
finite region in the parameter space. So we find that the generic  Hamiltonian 
within class DIII also possesses a stable gapless phase which is a WSM in two-dimension.  

This  WSM  is a topologically robust gapless phase.  We have confirmed the topological 
robustness of the WNs by calculating their topological  charges $W$. 
To find  the location of the WNs, let us assume $V$ is independent of momenta. Then the solution 
for WNs is given by the intersection of 
the circle $k_x^2 + k_y^2 = V^2$ with the conic section $ \sum_{ij} t^{(m)}_{ij} k_i k_j = m_0$. 
Since there is no linear term in momenta in the conic section, the curve has to be symmetric about
the origin $k_x=k_y=0$. Intersections between this circle and the  conic section, 
should occur at  four points (four WNs). We have numerically computed their 
topological charge and  find that the pair of WNs, which are time-reversal partners, carries 
same charges, which is  opposite to the charges carried by the other pair $W=1$. 

\begin{table}[ht]
  \begin{center}
  \caption{Comparing K-theory classification \cite{Zhao_Wang_2013, Matsuura_Chang_2013, Chiu_Schnyder_2014} of Weyl nodes in 2D  with that of the present approach. K-theory  
  predicts trivial index in class BDI and a $\mathbb{Z}_2$ index in class CII. A 
  $\mathbb{Z}_2$ index cannot protect Weyl nodes at off high-symmetry points in the BZ
  \cite{Chiu_Schnyder_2014}. Our results are consistent with the predictions in 
  Ref. \cite{Chiu_Ryu_2016} which later updated the K-theory classification.  }
    
    \begin{tabular}{ p{20mm}  p{15mm}  p{20mm} } 
      \hline 
      \hline
      \text{Class} & \text{K-theory} & \text{Our-approach} \\
      \hline
    
      AIII & $\mathbb{Z}$ & $\mathbb{Z}$  \\ 
     
      BDI  & 0 & $\mathbb{Z}$  \\ 
     
      DIII & $\mathbb{Z}$ & $\mathbb{Z}$ \\
      
      CII  & $\mathbb{Z}_2$ & $\mathbb{Z}$ \\
      
      CI   & $\mathbb{Z}$ & $\mathbb{Z}$  \\
      \hline 
    \end{tabular}
    \label{tab:result}
  \end{center}
\end{table}

\subsection{Discussion}

In summary of the foregoing analysis, we have demonstrated that the generic Hamiltonian 
within each chiral class in 2D exhibits a robust semimetal phase which is WSM in two-dimension. 
Notably, within the realm of 2D chiral classes, WSM stands as the sole stable 
semimetal phase. Importantly, the WSMs within these 2D chiral classes do not necessitate 
any crystalline symmetry (except for lattice translations) for the protection of Weyl points.
This result is also in accordance with the prediction in Ref. \cite{Chiu_Ryu_2016}.
Topological stability of this 2D WSM state stems from the nontrivial $\mathbb{Z}$ topological charge 
(see TABLE \ref{tab:result}) carried by the Weyl points. This topological charge, denoted as 
the winding number $W$ 
in Eq. \ref{eq:Winvariant}, serves as the invariant for 1D topological insulators in 
class AIII. Similar to WSMs in three dimensions, the sole means of annihilating a WN witin a 
symmetry class  is  by bringing two WNs of opposite topological charges into proximity and 
pairwise annihilation.

We have observed that in classes BDI and CII (CI and DIII), a  WN
carries a  topological charge which is opposite (identical) to its time-reversed partner. 
This occurrence is a consequence of the class-specific constraints: ${\cal T}^2 = {\cal C}^2$ 
in BDI and CII, compelling the topological charge $W$ of a WN to be opposite to its 
time-reversed partner's charge. Conversely, the constraint ${\cal T}^2 = -{\cal C}^2$ in 
CI and DIII results in WPs and their time-reversed partners possessing identical 
topological charges (see Appendix \ref{App:A} for a derivation). We will shortly see  
a consequence of this in a lattice model of WSMs in two dimensions.

Let us reconsider our choice to use a four-by-four Dirac Hamiltonian for all 
five chiral classes in two dimensions. While constructing the generic Hamiltonian 
for a chiral class, we imposed two key conditions: (i) the Hamiltonian must 
account for spin (denoted by $\sigma$), and (ii) it must exhibit chiral symmetry, 
acting on the sublattice space (denoted by $\eta$). This led to a four-by-four 
Hamiltonian. Additionally, for the sake of generality, we opted to start with 
a massive Dirac Hamiltonian. However, we note that this last condition is only 
essential for class DIII, which supports non-trivial topological gapped
states in two dimensions.

When considering massive nature of initial Dirac Hamiltonian, a notable 
constraint emerges for the two classes, CII and CI.
We observe that the Dirac Hamiltonian $H_0({\bf k})$ in class CII  (Eq. \ref{eq:HCII})  
and CI (Eq. \ref{eq:H0CI}) is inherently massless. 
In fact,  it is  not possible to add a  mass term to a four by four Dirac Hamiltonian 
belonging to CII and CI in two dimensions. 
The minimum matrix dimension required for a massive Dirac Hamiltonian within these classes 
is eight in two dimensions. This prompts an inquiry into whether the generic Hamiltonian 
in the chiral classes CII and CI, constructed from a more general eight-by-eight massive Dirac 
Hamiltonian, still exhibits the WSM state as a robust, gapless phase. 
The authors in the Refs. \cite{Abdulla_Das_2024} have constructed generic 
Hamiltonian from a eight-by-eight Dirac Hamiltonian by adding all the terms allowed 
in the symmetry classes CII and CI. Although,  the focus was in three spatial dimensions, 
their calculations extend to two dimensions with an appropriate modification of the kinetic 
term—specifically, the $k_z$ term is excluded. Subsequently, upon analyzing the solutions,
we observe the emergence of a WSM state as a robust gapless phase for the generic 
(eight-by-eight) Hamiltonian in both the  CII and CI classes.

We seek to juxtapose our findings with those of Refs. \cite{Shuichi_Murakami_2007, 
Burkov_Balents_2011, Abdulla_Das_2024}. In their works \cite{Shuichi_Murakami_2007, 
Burkov_Balents_2011}, it was established that the generic Hamiltonian  in the non chiral class 
AII in three  dimensions possesses a stable semimetal phase which exists as an intermediate 
phase between  the topological and trivial gapped states. This semimetal phase is a WSM 
in three dimensions. Recently, the Refs. \cite{Abdulla_Das_2024} demonstrated that the generic Hamiltonian in chiral classes AIII, DIII, CII and CI in three dimensions, possesses a 
stable nodal line semimetal phase which exists as an intermediate phase between  the 
topological and trivial gapped states, and these nodal lines are classified by 
the $\mathbb{Z}$ winding invariant (Eq. \ref{eq:Winvariant}). 
The presence of nodal lines in 3D and Weyl semimetals in 2D within the chiral classes and 
the fact that they are classified by the same $\mathbb{Z}$ winding invariant can  be 
attributed to the following  explanation: The topological classification of Fermi 
surfaces relies on both the co-dimension $p$ (=spatial dimension - Fermi surface 
dimension) of the Fermi surface \cite{Zhao_Wang_2013, Zhao_Wang_2014, Matsuura_Chang_2013, Chiu_Schnyder_2014, Chiu_Ryu_2016} and the symmetry of the 
restricted Hamiltonian used to calculate the topological invariant. Since both the WSMs in 2D 
and nodal line semimetals in 3D has same co-dimension $p = 2$, they are classified by the 
same $\mathbb{Z}$ winding invariant within the chiral classes. 

The presence of a stable semimetal phase within classes AII, AIII, DIII, CII, and CI in 3D can 
be intuitively grasped as follows. In three dimensions, each of the classes AII, AIII, DIII, CII, 
and CI accommodates a topologically nontrivial gapped state. The transition from a topologically nontrivial to a trivial gapped state must necessarily traverse through an intermediate gapless state. Consequently, the assurance of an intermediate gapless state's existence in a given  dimension is inherent to an AZ class exhibiting nontrivial topology in that dimension.

In view of the above argument,  our finding may not appear immediately apparent. With the 
exception of DIII, all other chiral classes exhibit trivial topology in two dimensions. However, 
we have demonstrated the existence of a stable semimetal phase within all five chiral classes 
in two dimensions. Remarkably, across all chiral classes, this gapless phase manifests as 
a WSM in the two-dimensional realm. This WSM state is topologically robust and does 
not require any crystalline symmetry for protection of the Weyl nodes.

\begin{figure}
\centering
\includegraphics[width=1\linewidth, height=0.48\linewidth]{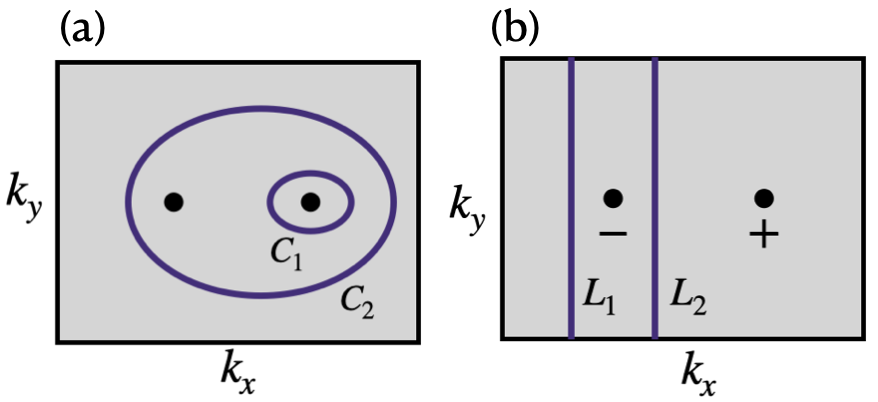}
\caption{(a) Two Weyl nodes (black dots) in a 2D BZ, with enclosing loops $C_1$, 
$C_2$.  (b) A pair of Weyl nodes with opposite  topological charges $W=\pm 1$. 
The vertical  lines $L_1$ and $L_2$ are two closed  lines at  a constant $k_x$ values.    }
\label{fig:fig1}
\end{figure}

%%%%%%%%%%%%%%%%%%%%%%%%%%%%%%%%%%%%%%%%%%%%%%

\section{Weyl semimetals and their boundary states in a lattice}
\label{Sec:FermiArc}

Having established the existence of  2D topological  WSM as a robust gapless phase 
in the continuum models within the chiral classes, now we want to write down a few 
illustrative lattice  models of  2D WSMs to study their topologically protected edge states. 
Before we move on to the discussion of edge states, we must  understand how  the 
topological charge $W$ transforms under time-reversal  in the different  chiral classes. 
In 3D WSM, the presence of additional time-reversal symmetry 
forces a WN to have monopole charge which is opposite to its time-reversal partner. Similarly, the 
presence of time-reversal and particle-hole symmetry in the BDI, CII, CI and DIII classes forces a 
WN to have a topological charge $W$ which is fixed by the charge of its time-reversed partner. 

As we have seen in the previous sections (for a proof see Appendix \ref{App:A}),  
a pair of WNs, which are related by 
time-reversal symmetry, carries opposite charges in class BDI and CII and identical 
charges in class CI and DIII. It is the relation ${\cal T}^2 = {\cal C}^2$ in class 
BDI and CII, which forces a WN to have charge $W$ which is opposite to its time-reversed
partner. Similarly it is ${\cal T}^2 = -{\cal C}^2$ in class CI and 
DIII, which puts condition on a WN to have charge $W$ which is identical to the charge 
carried by its  time-reversed partner.

Now we can easily establish from the periodicity of the BZ in a lattice  that the 
minimal number of WNs is two (four) in class BDI, CII (CI, DIII). Consider a 
loop $C_1$ which encloses a single WN. Its topological charge is given by the 
integration (in Eq.  \ref{eq:Winvariant}) over
the closed loop $C_1$. Now consider another closed loop $C_2$ which encloses all 
the  WNs in the 2D BZ (as in Fig. \ref{fig:fig1}(a)), then the total charge is given by 
\begin{align}\label{eq:Tcharge}
\sum_i W_i = \frac{1}{2\pi i} \int_{C_2} d{\bf k} \cdot \mathrm{Tr}\left(Q^{-1} \nabla_{\bf k} Q \right), 
\end{align}
where $W_i$'s are the charges of the individual WNs. Since there are no WNs (or gapless points)
outside the loop $C_2$, we can expand the loop $C_2$ continuously to the BZ boundary without 
altering the total charge $\sum_i W_i$. By periodicity of the BZ, the boundary of the BZ is equivalent to a point. This implies that the line integral in Eq. \ref{eq:Tcharge} must vanish. 
Therefore. the sum of the charges of all the WNs  must be zero in a lattice. We note  
that the above result for WNs in 2D is same as  the Nielsen-Ninomiya theorem \cite{Nielsen_Ninomiya_1983} for Weyl fermions in 3D in a lattice. Now it  is clear  
why the minimal number of WNs  has to be two (four) in class BDI,  CII  (CI, DIII). Since, 
there is no time-reversal or particle-hole  symmetry in class AIII,  the total  number  
of WPs in AIII models of WSMs will be a multiple of two. 

\subsection{Fermi arc edge states}

\begin{figure}
\centering
\includegraphics[width=0.7\linewidth]{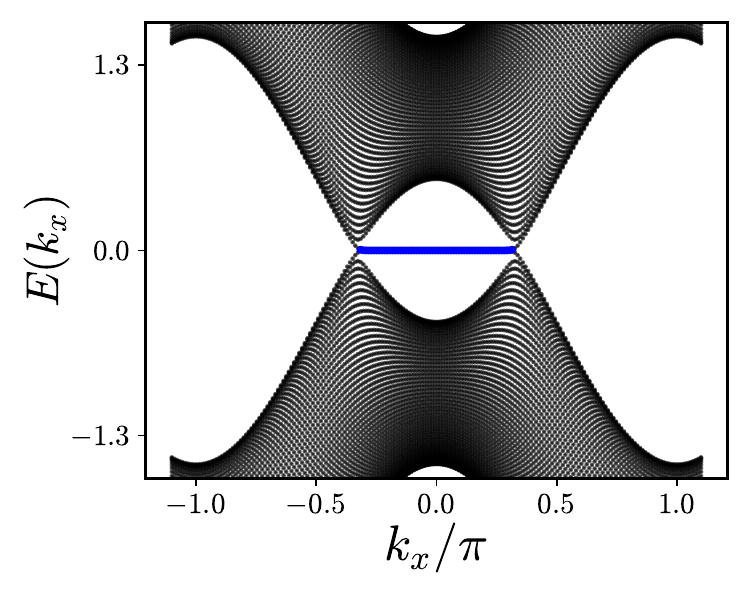}
\caption{Energy spectrum about zero energy of the BDI model with two 
WNs (Eq. \ref{eq:SimpleBDI}), in a cylindrical geometry with finite 
length $L_y=60$. The edge states (highlighted in blue), which exist only 
at the zero energy, forms an arc by  joining the projections of WNs of 
opposite charges  located at  $k_x/\pi = \pm 1/3$.   }
\label{fig:fig2}
\end{figure}

Existence of Fermi arc like flat surface states for point-node topological 
semimetals were previously discussed in the Refs. \cite{Matsuura_Chang_2013, 
Chiu_Schnyder_2014}. While Fermi arc surface states in 3D WSMs are well-understood,
we aim to apply similar reasoning to gain a clear understanding of edge states 
in 2D WSMs within the chiral classes. Additionally, we will demonstrate that 
the properties of edge states align with the $\mathbb{Z}$ classification of 
WNs within 2D chiral classes.

In 3D WSMs, the existence of surface states is  attributed to the nontrivial topological 
charge  (Chern number) carried  by  the Weyl nodes.  Consider a WSM with two WNs at 
${\bf k}=(k_0, 0, 0)$  and $-{\bf k}=
(-k_0, 0, 0)$. The Fermi arc surface states  are understood as a collection 
of the Chern insulator's (which exist for  fixed $k_x$ values in between $k_0$ and $-k_0$) 
chiral edge states \cite{Wan_Turner_2011, Burkov_Balents_2011, Burkov_Hook_2011}.
The above argument about the existence of Fermi arc surface states in 3D WSM  also 
applies to 2D WSMs within the chiral classes. The difference is that now we
have to consider 1D closed lines (instead of planes)  and the insulators living on the closed 
lines describe 1D AIII insulators (instead of Chern insulators) for the case of  WSMs in
2D chiral classes.

Let us briefly review the reasoning that 2D WSMs must exhibit 
Fermi arc edge states, analogous to the Fermi arc surface states in 3D Weyl semimetals. 
Consider two closed lines, $L_1$ and $L_2$, as depicted in Fig. \ref{fig:fig1}(b). 
Since the Hamiltonian, when restricted to arbitrary closed lines $L_1$ and $L_2$, 
preserves only chiral symmetry, it generally falls within class AIII. 
Suppose the Hamiltonian, which is restricted on 
the $L_1$, describes a trivial insulator, then the Hamiltonian restricted on 
$L_2$ must describe a 1D AIII topological insulator, because the line $L_2$ has crossed 
a WN of nonzero topological charge. Since  AIII topological  insulator in 1D has protected edge 
states, the 2D WSM must have topologically protected edge states which are given by the 
collection of  the 1D topological insulator's edge states  for all the $k_x$ values in between 
the two WNs of opposite topological charges. The zero energy edge states  must
lie on an arc (Fermi arc edge states) joining the projections of WNs of opposite charges 
on the 1D edge Brillouin zone. 

In 3D WSMs, the surface states are dispersionless along one direction, forming Fermi arc states, but disperse linearly along the transverse direction. In 2D WSMs, the reduced dimensionality eliminates the transverse direction, making the edge states entirely dispersionless. Due to the chiral symmetry protection, the edge states of 2D WSMs within the chiral classes are confined to zero energy, forming Fermi arcs that connect the projections of WNs with opposite charges 
on the 1D edge Brillouin zone.

Below we demonstrate the Fermi arc edge states of  2D WSMs  though a simple model, 
belonging to BDI class. Let us consider a two bands model with minimal two WNs, 
\begin{align}\label{eq:SimpleBDI}
H({\bf k}) = \left(1+\cos{k_0} - \cos{k_x} - \cos{k_y}\right) \eta_x  + \sin{k_y}\eta_y, 
\end{align} 
where $\eta$'s are the Pauli matrices which represent a pseudo spin. The Hamiltonian 
preserve time-reversal with ${\cal T} = {\cal K}$ and particle-hole with ${\cal C} = \eta_z {\cal K}$. Since ${\cal T}^2 = {\cal C}^2 =1$, the Hamiltonian in Eq. \ref{eq:SimpleBDI} belongs 
to class BDI. There are only two  WNs located at ${\bf k}_w = (\pm k_0, 0)$. Computing their topological charges using Eq. \ref{eq:Winvariant}, we find  $W = \pm 1$ respectively. 

To observe the edge states associated with the WNs, the system must be finite and 
open in a suitable direction. As the two WNs have non-overlapping projections 
only in the 1D $k_x$ edge BZ, the system must be finite along the $y$ direction to 
reveal its topological Fermi arc edge states. We calculated the energy spectrum by
exact diagonalization in a cylindrical geometry with a finite length $L_y=60$. 
Figure \ref{fig:fig2} presents the relevant energy spectrum near zero energy, 
highlighting the edge states. The zero-energy edge states (all edge states are 
at zero energy) form a Fermi arc connecting the projections of WNs with opposite 
charges.

\begin{figure}
\centering
\includegraphics[width=1\linewidth]{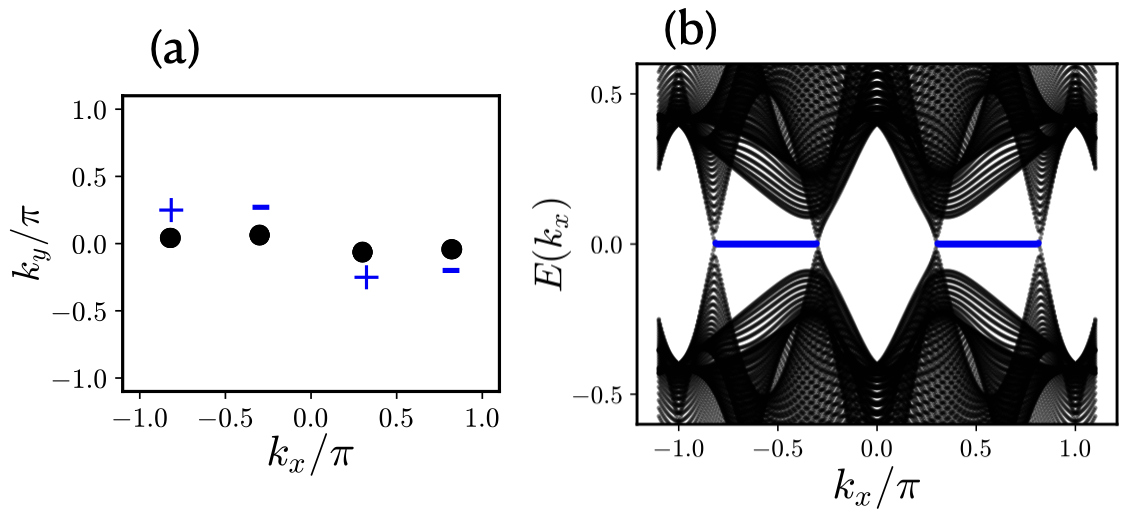}
\caption{(a) Location of the WNs of the  CII model in Eq. \ref{eq:HLCII} and 
the sign of their topological charges. 
(b) Energy spectrum (about zero energy) in a cylindrical  geometry with finite 
length $L_y=80$.  The edge states (highlighted in blue), which exist only  at 
the zero energy, forms a Fermi arc by  joining the projections of WNs of 
opposite charges $W=\pm 1$. }
\label{fig:fig3}
\end{figure}

\subsubsection{Fermi arc edge states in CI and CII models}

A  straightforward way to obtain lattice model is to take the continuum model
of 2D WSMs (as given in Eqs. \ref{eq:HAIII},  \ref{eq:HBDI},  \ref{eq:HCII},  \ref{eq:HCI} 
and  \ref{eq:HDIII}) and replace  (i) the linear terms in ``$k_i$'' by 
``$\sin{k_i}$'' and ii) the quadratic terms   by ``$\cos{k_i}$''. 
Following this prescription, we can write down a general lattice (hopping) model
\begin{align}\label{eq:LatticeModel}
    H = \sum_{{\bf n}, j} c^{\dagger}({\bf n})~ 2 {\cal M} ~ c({\bf n}) - \left(c^{\dagger}({\bf n} + a \hat{e}_j) 
    ~ {\cal T}_j ~ c({\bf n}) + H.c \right), 
\end{align}
which is defined on a square lattice. Here $n = a(n_1,n_2)$, $n_i$ being integers, denotes
the lattice sites, $\hat{e}_j$
is the unit vector along $j^{th}$ direction. We will set the lattice constant $a=1$.
The matrix dimension of the onsite term ${\cal M}$ and the hopping term ${\cal T}_j$ (should not 
be confused with the time-reversal operator ${\cal T}$ which does not have subscript)
depends on the symmetry class, and it is given by the matrix dimension of the Dirac 
Hamiltonian in the symmetry classes. Note that the onsite term ${\cal M}$ includes all the 
momentum-independent terms (as given 
in the continuum Dirac models)  allowed by the symmetry class. The hopping matrices 
${\cal T}_j$, $j=1, 2$, may be expressed in
a generic form ${\cal T}_j = t_j  \tilde{\gamma} + i\gamma_j $, where $i$ is the imaginary 
unit and $t_j$'s are real numbers. The $\gamma_j$ matrices, $j=1, 2$, anticommutes \
with each other and $\tilde{\gamma}$  need not to anticommute with $\gamma_j$ (as is 
the case for  four bands models in class CI and CII). Now we can explicitly specify $\gamma_j, \tilde{\gamma},  {\cal M}$ and  ${\cal T}_j$  to obtain a lattice 
model of 2D WSM in each of the five chiral classes and investigate  their Fermi arc 
edge states.

We have seen earlier that the Hamiltonian in class  BDI, CII obey ${\cal T}^2=
{\cal C}^2$ and those in class CI, DIII obey  ${\cal T}^2=-{\cal C}^2$, and  a WN  
carries a topological charge which is opposite (equal) to the charge carried by its 
time-reversed partner in class BDI, CII (CI, DIII) in two dimensions. It is needless 
to go through all the chiral classes, here we choose two chiral class: one  with  
${\cal T}^2={\cal C}^2$ and the other with ${\cal T}^2=-{\cal C}^2$ to demonstrate 
the existence of Fermi arc edge states associated  with the Weyl nodes. 

In class  CII, the Dirac Hamiltonian in Eq. \ref{eq:HCII} is four by four, 
therefore $c^{\dagger}({\bf n})$ and $c({\bf n})$ are four components Dirac 
fermion creation and annihilation operators respectively.
From the Dirac Hamiltonian in Eq. \ref{eq:HCII}, we read the two anticommuting Gamma matrices 
$\gamma_1 = \eta_x \sigma_x$ and $\gamma_2=\eta_x \sigma_z$. We choose $\tilde{\gamma} = 
\eta_x \sigma_0$, which is the $V_1$ term in Eq. \ref{eq:HCII}. Then the matrix ${\cal M}$ which 
contains the momentum independent terms of $H_{\textrm{CII}}$  is 
\begin{align*}
{\cal M} =  V_1 \eta_x \sigma_0 + V_2 \eta_y \sigma_y +   {\bf A} \cdot {\boldsymbol \alpha}. 
\end{align*}
Therefore, a CII lattice model (Fourier transforming $H$ in Eq. \ref{eq:LatticeModel}) of WSM in 
two dimensions is given by 
\begin{align} \label{eq:HLCII}
H_{\textrm{CII, lat}}({\bf k}) =  &   \sin{k_x} \cdot \eta_x \sigma_x +   
\sin{k_y} \cdot \eta_x \sigma_z   \nonumber  \\  
& + \left({\cal M} -  (t_1 \cos{k_x} + t_2 \cos{k_y})\eta_x \right).
\end{align}
For  the Femi arc edge states associated with the WPs,  we take a representative 
choice of the parameters: $t_1=0.7, t_2=0.8$, $V_1=0.5, V_2=0.2$ and ${\bf A} = (0.1, 0.4)$. 
For this choice of parameters, there exist four WNs as shown in Fig. \ref{fig:fig3}(a). Clearly,
their projections on the $k_x$ axis are non-overlapping and they are well separated. 
Therefore the Fermi arc edge states should be clearly seen when the system is taken finite 
with open boundary along the $y$-direction so that $k_x$ remains a good quantum number. 
We have numerically computed the spectrum of the lattice model in Eq. \ref{eq:HLCII}, 
in a cylindrical geometry with finite length $L_y=80$ along the $y$-direction and 
plotted it in Fig. \ref{fig:fig3}(b). The zero energy edge states, which lie in between 
the bulk valence and conduction  bands,  join the  projections of WNs of opposite 
charges to form an arc. 

\begin{figure}
\centering
\includegraphics[width=1\linewidth]{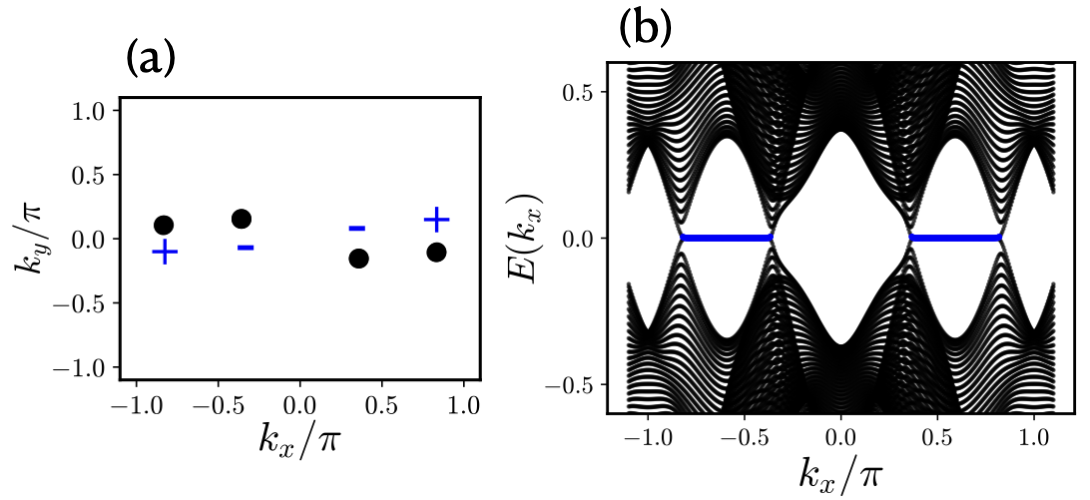}
\caption{(a) Location of the WNs in the CI model from Eq. \ref{eq:HLCI}. 
Note that the pair of WNs, related by time-reversal symmetry, carry the 
same charge. (b) Energy spectrum (near zero energy) in a cylindrical
geometry with a finite length of $L_y = 80$. The edge states (highlighted 
in blue), existing solely at zero energy, form a Fermi arc connecting the 
projections of WNs with opposite charges $W=\pm 1$.  }
\label{fig:fig4}
\end{figure}

For class CI, the two anticommuting Gamma matrices can 
be easily read  from the Dirac Hamiltonian in Eq. \ref{eq:HCI}: $\gamma_1= 
\eta_x \sigma_y$ and $\gamma_2=\eta_y\sigma_y$.  We can  choose $\tilde{\gamma} = 
\eta_x \sigma_0$, which is the $V_1$ term in Eq. \ref{eq:HCI}. Then  the matrix 
${\cal M}$ which  contains the momentum independent terms 
of $H_{\textrm{CI}}$ in Eq. \ref{eq:HCI} is 
\begin{align*}
{\cal M} = V_1 \eta_x +  V_2 \eta_y  +  {\bf A} \cdot {\boldsymbol \alpha} + {\bf B} \cdot {\boldsymbol \beta},  
\end{align*}
where ${\boldsymbol \alpha}$, ${\boldsymbol \beta}$ have been defined in Eq. \ref{eq:HCI}. 
Therefore, a CI lattice model (Fourier transforming $H$ in Eq. \ref{eq:LatticeModel}) of WSM 
is given by 
\begin{align} \label{eq:HLCI}
H_{\textrm{CI, lat}}({\bf k}) =  &  \sin{k_x}\cdot \eta_x \sigma_y +   \sin{k_y} \cdot \eta_y \sigma_y   \nonumber  \\  
& + \left({\cal M} -  (t_1 \cos{k_x} + t_2 \cos{k_y})\eta_x \right). 
\end{align}
We take a representative choice of parameters: $t_1=0.7, t_2=0.9$, $V_1=0.4, 
V_2=0.3$,  ${\bf A} = (0.3, 0.2)$  and ${\bf B} = (0.5, 0.3)$, so that the Hamiltonian in 
Eq. \ref{eq:HLCI} describes a WSM state. For this choice of parameters, the WSM state 
hosts four WNs as depicted in Fig. \ref{fig:fig4}(a). For the 
Fermi arc edge states,  we take the system finite along the $y$-direction and compute 
the spectrum numerically. The spectrum around zero energy is  plotted in 
Fig. \ref{fig:fig4}(b). Similar to the previous cases of edge states in 
class BDI and CII, the  edge states in CI also exist only at zero energy, forming 
an arc that connects the projections of WNs of opposite topological 
charges.

In the preceding, we derived lattice models from the continuum Dirac models
in a straightforward way by substituting the linear terms 
in ``$k_i$''  with ``$\sin{k_i}$" and the 
quadratic terms with ``$\cos{k_i}$",  without delving into the specifics of the 
lattice structure and the sources of these terms. Now, as we move forward, we 
turn our attention to more  realistic systems, namely Graphene and a 2D SSH model, 
to demonstrate that when chiral symmetry is maintained, WNs consistently emerge 
within their gapless phases.

\subsection{Graphene and 2D SSH model}

Graphene is a topological semimetal with Dirac cones at the Fermi energy. The Bloch 
Hamiltonian \cite{Konschuh_Fabian_2010, Kochan_Fabian_2017}, for a nearest neighbour 
hopping model, is 
\begin{align} \label{eq:GrapheneBlochH0}
H({\bf k}) =  \begin{pmatrix}
0 & f({\bf k}) \\
f^{*}({\bf k}) & 0  
\end{pmatrix}   
 =  f_1({\bf k}) \sigma_x + f_2({\bf k}) \sigma_y, 
\end{align}
where $f({\bf k}) = 1 + e^{i{\bf k} \cdot {\bf R}_2} +  e^{-i{\bf k} \cdot {\bf R}_3}$ and 
$f_1({\bf k}) = 1+ \cos{{\bf k} \cdot {\bf R}_2} +  \cos{{\bf k} \cdot {\bf R}_3}$,  $f_2({\bf k}) 
=  -\sin{{\bf k} \cdot {\bf R}_2} +  \sin{{\bf k} \cdot {\bf R}_3}$. The lattice vectors ${\bf R}_n$ 
($n=1,2,3$) are given by ${\bf R}_n = a\left(\cos{2\pi(n-1)/3}, ~\sin{2\pi(n-1)/3}\right)$. 
Here $\sigma$'s represent the 
two sublattices  A and B of the graphene honeycomb lattice. The spinless graphene Hamiltonian 
in Eq. \ref{eq:GrapheneBlochH0} obey time-reversal with ${\cal T} = {\cal K}$ and particle-hole 
with ${\cal C} = \sigma_z {\cal K}$. The Hamiltonian has chiral symmetry ${\cal S} = \sigma_z$ 
which originates due to the sublattice symmetry. Clearly  time-reversal and particle-hole symmetry
obey ${\cal T}^2={\cal C}^2=1$. Therefore the spinless graphene Hamiltonian in Eq. \ref{eq:GrapheneBlochH0}
belongs to the chiral class BDI. There are two gapless points in the energy spectrum at 
$\pm {\bf K} = \pm \frac{4\pi}{3a}(1, 0)$ and the low energy spectrums around $\pm {\bf K}$ form 
Dirac cone. For spinless graphene, the Dirac cones are nondegenerate. Therefore, according to 
the definition of WNs in 2D (discussed in the introduction),  the touching points $\pm {\bf K} $ 
of spinless graphene are called Weyl nodes. Recall that a WSM  belonging to the BDI class 
possesses a minimum of two WNs carrying opposite topological charges, and  graphene satisfies 
this requirement.  To verify topological protection of the  WNs, we compute their 
topological charge $W$, using Eq. \ref{eq:Winvariant}.  We find that they carry 
opposite topological charges $W=\pm 1$. Thus, the spinless graphene (described by Eq. \ref{eq:GrapheneBlochH0}) is a 2D  WSM  in  class BDI. 

As we have discussed earlier,  the WNs in spinless graphene  
cannot be removed by small perturbations (provided they do not alter 
the symmetry class) unless  the two  WNs come  close   and  annihilate 
each other pairwise. However, 
if the chiral symmetry broken  perturbations are added, the  WNs loose their 
topological protection and can get  gapped out immediately.  
One illustrative instance of such perturbation is the inclusion of a spin-orbit 
coupling term \cite{Kochan_Fabian_2017}
\begin{align}
h_{\lambda}({\bf k}) = \lambda g({\bf k}) \sigma_z \otimes s_z, 
\end{align}
where $g({\bf k})  = -\frac{2}{3\sqrt{3}} \sum_{n=1,2,3} \sin{{\bf k} \cdot {\bf R}_n}$, and 
the Pauli matrix $s_z$ represents  electron's spin. This term breaks  the chiral symmetry 
because of the  $\sigma_z$ dependence in $h_{\lambda}({\bf k})$, immediately creates  a 
gap in the spectrum.

\begin{figure}
\centering
\includegraphics[width=0.6\linewidth, height=0.5\linewidth]{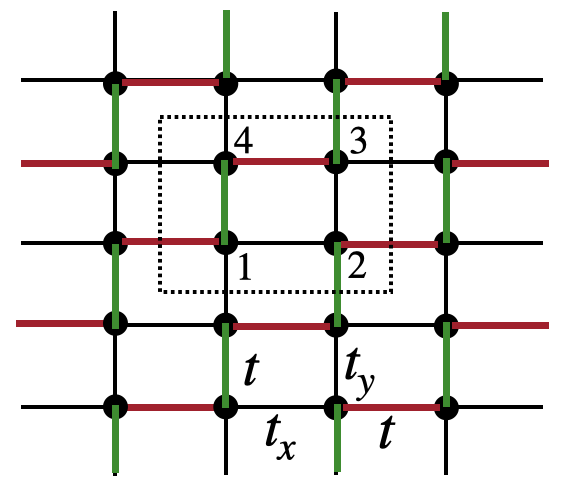}
\caption{Schematic of 2D lattice where SSH model is defined. The dimerized hopping 
along $x$-direction is denoted by $t$, $t_x$  and those along $y$-direction are 
$t$, $t_y$.  }
\label{fig:fig5}
\end{figure}

Let us now examine another  example of  WSM protected by chiral/sublattice symmetry  
in two-dimension: In a 2D SSH (hopping) model \cite{Zhang_Zhou_2017_two, Li_Song_2017, Li_Trauzettel_2022} featuring dimerized  bonds along both the $x$ and $y$ directions, 
as depicted in  Fig. \ref{fig:fig5}, we can represent it as follows 
\begin{align}\label{eq:HSSH}
H =  \sum_{{\bf R}} & \left(t_x d^{\dagger}_1({\bf R}) d_2({\bf R}) + t d^{\dagger}_4({\bf R}) d_3({\bf R}) + \textrm{H.c} \right) \nonumber \\
&   + \left(t_y d^{\dagger}_2({\bf R}) d_3({\bf R}) + t d^{\dagger}_1({\bf R}) d_4({\bf R}) + \textrm{H.c} \right) \nonumber \\
& +  \left(t d^{\dagger}_2({\bf R}) d_1({\bf R} + \hat{x}) + t_x d^{\dagger}_3({\bf R}) d_4({\bf R}+\hat{x}) + \textrm{H.c} \right) \nonumber \\
& +  \left(t_y d^{\dagger}_4({\bf R}) d_1({\bf R} + \hat{y}) + t d^{\dagger}_3({\bf R}) d_2({\bf R}+\hat{y}) + \textrm{H.c} \right).\nonumber \\
\end{align}
There are four lattice sites, labelled $\alpha=1, 2, 3, 4$, in the unit cell. 
To diagonolize $H$, we Fourier transform $d^{\dagger}_{\alpha}({\bf R})$ and $d_{\alpha}({\bf R})$ to get the following  Bloch Hamiltonian in the basis $d({\bf k}) = \left(d_1({\bf k}) ~ d_3({\bf k}) ~ d_2({\bf k}) ~ d_4({\bf k})\right)^T$
\begin{align} \label{eq:HKSSH}
H({\bf k}) = \begin{pmatrix}
0 & Q({\bf k}) \\
Q^{\dagger}({\bf k}) & 0 \\
\end{pmatrix},
\end{align}
where $Q({\bf k})$ is a two by two matrix 
\begin{align}
Q({\bf k}) =  \begin{pmatrix}
t_x + te^{ik_x}  & t+ t_y e^{ik_y} \\
t_y + te^{-ik_y} & t+ t_x e^{-ik_x} \\
\end{pmatrix}.
\end{align}
For symmetry analysis, it is useful to rewrite the Bloch Hamiltonian 
in Eq. \ref{eq:HKSSH} in the following compact form
\begin{align}
H({\bf k}) = &  f_{x1} \cdot \eta_x \tau_0 + f_{x2}\cdot \eta_y \tau_0 +   f_{x3} \cdot \eta_x \tau_z +  
f_{x4} \cdot \eta_y \tau_z \nonumber \\
& +  f_{y1} \cdot \eta_x \tau_x + f_{y2}\cdot \eta_y \tau_x +   f_{y3} \cdot \eta_x \tau_y + 
 f_{y4} \cdot \eta_y \tau_y, 
\end{align}
where $f_{x1}({\bf k}) = (t+t_x)(1+\cos{k_x})/2$, $f_{x2}({\bf k}) = (t_x-t)\sin{k_x}/2$, 
$f_{x3}({\bf k}) = (t_x-t)(1-\cos{k_x})/2$, $f_{x4}({\bf k}) = -(t_x+t)\sin{k_x}/2$, $f_{y1}({\bf k}) 
= (t_y+t)(1+\cos{k_y})/2$,
$f_{y2}({\bf k}) = (t-t_y)\sin{k_y}/2$, $f_{y3}({\bf k}) = (t_y+t)\sin{k_y}/2$, $f_{y4}({\bf k}) 
= (t_y-t)(1-\cos{k_y})/2$.  We notice that the function $f_{xi/yi}({\bf k})$  multiplied to 
$\eta_{\mu} \tau_{\nu}$ is odd (even), when $\eta_{\mu} \tau_{\nu}$ is complex (real). 
Therefore the Hamiltonian $H({\bf k})$ in Eq. \ref{eq:HKSSH} is  symmetric under 
time-reversal with ${\cal T} = {\cal K}$. It also obeys particle-hole symmetry with  ${\cal C} 
= \eta_z {\cal K}$. The chiral symmetry ${\cal S} = \eta_z$ is obvious from the 
block-offdiagonal form of $H({\bf k})$ in Eq. \ref{eq:HKSSH}. Since ${\cal T}^2 = 
{\cal C}^2=1$, the 2D SSH Hamiltonian $H$ in  Eq. \ref{eq:HSSH} belongs to the 
chiral class BDI. 

In the following, we will see that the 2D SSH model in Eq. \ref{eq:HSSH} possesses a 
stable gapless phase (in its phase diagram) which is a  topological WSM in
two-dimension. To find gapless phases, we need to compute the zeros of the 
energy spectrum $E({\bf k})$. The  energy spectrum  is given by
\begin{align}
& E^2_{\xi} ({\bf k}) = A_{\xi}^2({\bf k}) + B_{\xi}^2({\bf k}), \nonumber \\
& A_{\xi}({\bf k})  = (t+t_x)\cos{\frac{k_x}{2}} + \xi(t+t_y)\cos{\frac{k_y}{2}}, \nonumber \\
& B_{\xi}({\bf k})  = (t-t_x)\sin{\frac{k_x}{2}} - \xi(t-t_y)\sin{\frac{k_y}{2}}, \nonumber \\ 
\end{align}
where $\xi = \pm 1$. The gapless points, which are given by the intersection of the two 
smooth curves   $A_{\xi}({\bf k}) = 0$ and  $B_{\xi}({\bf k})=0$, are generically Weyl nodes. 
They are located at $\pm {\bf K} = \pm (K_x, - K_y)$, where $K_{x/y}$ 
is given by
\begin{align}
K_{x/y} = 2 \cos^{-1}\left(\sqrt{ \frac{(t + t_{y/x})^2 (2t -t_x -t_y)}{4t(t^2 - t_x t_y)} }\right).
\end{align}
As long as,  the quantity inside the square root  lies in the range $[0, 1)$, small change 
in the parameters $t, t_x, t_y$ can not remove these Weyl nodes. There is a finite region 
in the parameter space  for solution of Weyl nodes. The pair of WNs are topologically 
protected because they carry nonzero and opposite topological charges $W=\pm 1$. Any 
small  perturbation, within the class BDI,   cannot gap out  the WNs unless they 
meet at some point in the BZ and annihilate  each other.

%%%%%%%%%%%%%%%%%%%%%%%%%%%%%%%%%%%%%%%%%%%%

\section{Discussion and Conclusion}
\label{Sec:Summary}

% Previously, the WSMs in two dimension have been believed to require some form of 
% crystalline symmetry (e.g. inversion, mirror, rotation, nonsymmorphic symmetries), 
% in addition to the internal symmetries, to exist as a robust  phase. 

% Depending on which crystalline symmetries are present, the classification of the 
% WPs gets modified  and so the topological charges they carry. Furthermore, the 
% boundary states which are 
% considered as hallmark of nontrivial bulk topology, may not be protected because 
% the protecting crystalline symmetries could be in general broken for a finite size 
% systems. 

In this work,  we have unambiguously shown that the generic Hamiltonian within 
the 2D chiral classes possesses  a robust semimetal  phase which is  a topological 
Weyl semimetal. Notably, within the realm of 2D chiral classes, WSM stands 
as the sole stable gapless semimetal phase. Most importantly, this WSM state within 
the chiral classes does not require any crystalline symmetry for its protection. 
Independent to which chiral class, the WSM state belongs to, the state is 
characterized by a single $\mathbb{Z}$ invariant. 
%which is in accordance with 
%the prediction in Ref. \cite{Chiu_Ryu_2016}. 

Similar to the Fermi arc surface states of WSM in 3D,  the 2D WSMs in the chiral class
also host Fermi arc edge  states which always  join the projections of WNs
of opposite topological charges on the edge Brillouin zone.  However there is an important 
distinction between the Fermi arc surface states in 3D WSMs and Fermi arc edge states 
in 2D WSMs: In  case of 3D WSMs, the surface states exhibit no dispersion 
in one direction, but they disperse linearly in the perpendicular direction. Conversely, in 
2D WSMs,  edge states are completely dispersionless due  to the  reduced 
dimensionality. 

The topological electromagnetic response of topological insulators/semimetals/metals 
serves as a crucial tool for comprehending the relationship between various physical 
quantities and the nontrivial topology present in their band structures. The 
correlation between physical quantities and topological invariants is well-established in 
non-chiral classes \cite{TKNN_1982, Kane_Mele_2005, Qi_Zhang_2008}. In  case of a WSM in 3D, 
the topological charge of a Weyl node, expressed by the Chern number obtained through 
integrating the Berry curvature over a closed 2D surface surrounding the node, is linked 
to the Hall conductivity. The presence of nontrivial Berry curvatures around Weyl nodes 
and their associated Chern numbers results in diverse topological responses, such as 
anomalous Hall conductance, negative magneto-resistance, and planar Hall effect 
\cite{Aji_2012, Zyuzin_Burkov_2012, Son_Spivak_2013, Gorbar_Miransky_2014, Burkov_2015, 
Li_Roy_2016, Lu_Shen_2017, Nandy_Tewari_2017, Li_Shen_2018, Shama_Singh_2020, 
Li_Yao_2023}. 
Conversely, for chiral classes, the relationship between the topological invariant (winding 
number) and physical quantities is less straightforward. However, if the chiral symmetry 
is weakly broken \cite{Hosur_Vishwanath_2010, Shiozaki_2013, Wang_Duan_2015}, a
 finite quantized Hall response is anticipated. Given this electromagnetic 
response of topological insulators characterized by winding numbers within the chiral class, 
we anticipate a measurable electromagnetic response (Hall conductance) for WSMs 
within the 2D chiral classes, provided a chiral symmetry-broken perturbation 
is introduced to open a  finite gap in the system. Recently, (001) oriented 
Cd$_3$As$_2$ thin film under an appropriate in-plane magnetic field \cite{Stemmer_2023, Stemmer_2024, Smith_2024} is reported to be a WSM in two dimensions, which exhibits 
quantized Hall conductance when a small perpendicular magnetic field is applied to open a 
finite gap in the system.

Upon incorporating our findings with those  in the Refs. \cite{Shuichi_Murakami_2007, 
Burkov_Balents_2011, Sato_Masatoshi_2006, Beri_2010, Abdulla_Das_2024}, a noteworthy 
relationship 
becomes evident—linking an AZ symmetry class to the topological semimetal (protected by 
internal symmetry) it harbors as a robust gapless phase:
The generic Hamiltonian belonging to  the non chiral classes (e.g. AII, A, AI) 
possesses WSM as a robust semimetallic phase in three dimensions. 
Nodal line semimetals in 3D protected by the internal symmetries  exist as a robust gapless 
phase within the 3D chiral classes. However within the 2D chiral classes,  WSMs  emerge 
as a  robust topological semimetal protected  by internal symmetries. Given this interesting 
observation, it is natural to ask whether there exists a robust topological semimetal 
protected by internal symmetry within the non chiral classes in two dimensions. 
K-theory classification \cite{Zhao_Wang_2013, Matsuura_Chang_2013, Zhao_Wang_2014, Chiu_Schnyder_2014, Chiu_Ryu_2016} predicts nodal line. In the near future, we hope 
to explicitly show its existence as a generic gapless phase within the 2D non chriral 
classes.

Many other open questions remain. An important one is to pursue whether a WSM state in the 
2D chiral classes 
survive in presence of disorder. In case of WSM in 3D, it is known that the WSM state 
is perturbatively stable to disorder \cite{Fradkin_1986a, Fradkin_1986b, Goswami_Chakravarty_2011, 
Altland_Bagrets_2015, Kobayashi_Herbut_2014, Sbierski_Brouwer_2014, Louvet_Fedorenko_2016, 
Roy_RobertJan_2018}, where  the WSM undergoes a transition to a diffusive
metal at a finite critical disorder strength. There is another line of argument where 
the rare region  effects have been believed to destroy the WSM at arbitrarily weak 
disorder \cite{Nandkishore_Sondhi_2014, Pixley_Sarma_2016a, Pixley_Sarma_2016b, 
Pixley_Sarma_2017}. 
However in case of WSM state in the 2D chiral classes, a generic disorder which 
may break  the protecting chiral symmetry is expected to  destroy the WSM state  
by gapping out  the Weyl nodes. An interesting future direction would involve  
investigating the fate of the WSM state within 2D chiral classes,  in presence of  
chiral symmetric disorder (off-diagonal disorders).

Recent research, as highlighted in Refs. \cite{Abdulla_2023, Abdulla_Murthy_2022}, 
has showcased the profound implications of pairwise annihilation in a three 
dimensional WSMs, unveiling a myriad of 
novel phases in the presence of an external orbital magnetic fields. 
We foresee the emergence of similarly compelling physics in the context of WSMs in 
two dimensions. 

In conclusion, we emphasize that our approach is quite general and can 
readily be extended to various types of Fermi surfaces (or nodal surfaces) 
across different dimensions. For Fermi surfaces (or nodal surfaces) of 
co-dimension  $p$ , located away from high-symmetry points within the BZ
in the AZ class, the Hamiltonian restricted to an enclosing surface generally 
belongs to class A or AIII. Consequently, we expect these surfaces can be 
classified solely by $\mathbb{Z}$  invariants. Fermi surfaces of co-dimension  
$p$  within nonchiral (chiral) classes are classified by a  $(p-1)$ -dimensional 
class A (AIII)  Z  invariant. This prediction aligns with Refs. 
\cite{Chiu_Schnyder_2014, Chiu_Ryu_2016}, which argued that  $\mathbb{Z}_2$  
invariants, as predicted by K-theory \cite{Zhao_Wang_2013, Matsuura_Chang_2013, Zhao_Wang_2014}, 
cannot protect Fermi surfaces (or nodal surfaces) located away from 
high-symmetry points in the Brillouin zone.

%%%%%%%%%%%%%%%%%%%%%%%%%%%%%%%%%%%%%%%%%%%%%%%
\begin{acknowledgements}

The author wants to thank Ganpathy Murthy and Ankur Das for insightful  discussion. 
I would like to thank Sumathi Rao for her valuable comments on the manuscript. 
The author acknowledges the financial support provided by the Infosys Foundation. 
Additionally, I wish to express my sincere thanks to the Internation Centre for 
Theoretical Science (ICTS) for their warm hospitality during my visit, where a 
significant portion of this work was undertaken.

\end{acknowledgements}

%%%%%%%%%%%%%%%%%%%%%%%%%%%%%%%%%%%%%%%%%%%%%%%

\appendix

\section{Transformation of $W$ under time-reversal in different chiral classes}
\label{App:A}

\begin{figure}
\centering
\includegraphics[width=0.7\linewidth]{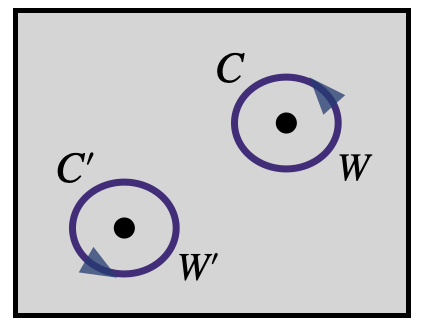}
\caption{Two Weyl npdes, which are time-reversed partner of each other at ${\bf k}_0$ and 
$-{\bf k}_0$,  carry  topological charges $W$ and $W'$ respectively.  Under time-reversal, 
the integration loop $C$ goes  to $C'$ without altering the orientation i.e. it remains 
counterclockwise.  }
\label{fig:fig6}
\end{figure}

In the main text, we have seen that  a WN carries a topological  charge $W$ 
(defined in Eq. \ref{eq:Winvariant})  which is opposite (equal) to the charge carried by 
its time-reversed partner in class BDI, CII (DIII, CI). In this appendix, we 
provide a proof of this. For convenience, let us work with a basis in which the chirality 
operator ${\cal S} = \eta_z \otimes {\mathbb I}_n$ (${\mathbb I}_n$ is a $n\times n$ identity 
matrix, $\eta$'s are Pauli matrices) is diagonal. The Bloch-Hamiltonian is  a $2n \times 2n$ 
dimensional matrix. 
Given the chirality operator ${\cal S} = \eta_z \otimes {\mathbb I}_n$ , we can choose the 
time-reversal operator to be  (i) ${\cal T} 
= i \eta_0 U {\cal K}$ or (ii) ${\cal T} = i \eta_x \tilde{U} {\cal K}$, where $U$, $\tilde{U}$  are
$n\times n$  unitary matrices satisfying $U U^{*} = \alpha$, $\tilde{U} \tilde{U}^{*} = \alpha$, 
with $\alpha=\pm 1$.  Note that we have an alternative choice (i) ${\cal T}  = i \eta_z U {\cal K}$ 
or (ii) ${\cal T} = i \eta_y \tilde{U} {\cal K}$, but they are identical to the former choice  upto 
a unitary transformation in the $\eta$ space. Without loss of generality, we will work with 
the former choice. Now from ${\cal S} = {\cal T C}$, it follows that (i) ${\cal C} = i \eta_z U {\cal K}$ 
or (ii) ${\cal C} = i \eta_y {\tilde U} {\cal K}$. Therefore, in the basis in which ${\cal S} = \eta_z {\mathbb I}_n$, 
the time-reversal and particle-hole operators are given by either 
\begin{align}
\textrm{(i)} ~~ {\cal T}  = i \eta_0 U {\cal K}, ~~ {\cal C} = i \eta_z U {\cal K},  ~~ U^{*}U = \alpha, 
\end{align}
or 
\begin{align}
\textrm{(ii)} ~~ {\cal T}  = i \eta_x \tilde{U} {\cal K}, ~~ {\cal C} = i \eta_y \tilde{U} {\cal K},  ~~  \tilde{U}^{*}\tilde{U} = \alpha. 
\end{align}
The case (i), in which $ {\cal T} ^2 =  {\cal C}^2 = \alpha$, represents the symmetry class 
BDI, if  $\alpha=1$, and  CII, if $\alpha=-1$.  And the case (ii), in which $ {\cal T} ^2 = 
-{\cal C}^2 = \alpha$, represent 
the DIII (for $\alpha=-1$) or CI (for $\alpha=1$) symmetry class. Now we are ready to show how the 
topological charge of a WN transform under  time-reversal  within the BDI, CII,  DIII, CI classes. 
To begin, we will concentrate on the BDI and CII classes. Suppose we have a pair of WNs at ${\bf k_0}$
and  at time-reversed momentum $-{\bf k_0}$ which carry charges $W$ and $W'$ respectively. 
We want to find a relation between 
$W$ and $W'$. The charge $W$ carried by the WN at ${\bf k}_0$ is given by the line-integral in 
 Eq. \ref{eq:Winvariant} over the close loop $C$ (which is traversed counterclockwise) as shown 
 in Fig. \ref{fig:fig6}. Under time-reversal, the topological charge $W$ goes to 
\begin{equation}
 \begin{aligned}\label{eq:Wpinvariant}
 W & \to  \frac{1}{2\pi i} \int_{C'} d{\bf k} \cdot \mathrm{Tr}\left({Q'}^{-1} \nabla_{\bf k} Q' \right)  \\
 & =  \frac{1}{2\pi } \int_{C'} d{\bf k} \cdot \textrm{Im}\left[\mathrm{Tr}\left({Q'}^{-1} \nabla_{\bf k} Q' \right) \right]
 \end{aligned}
 \end{equation}
which must be equal to the topological charge $W'$ of the time-reversed WN at $-{\bf k}_0$. 
In going from first to second line, we have used the fact that  $W$ is real to
rewrite the integral in a form which will be useful later.  To relate $W'$ with $W$, we have to 
find how $Q({\bf k})$ transforms  under time-reversal i.e.  we have to relate $Q'({\bf k})$ with  
$Q({\bf k})$. 

The $Q'({\bf k})$ can be obtained by 
evaluating ${\cal T} H({\bf k}) {\cal T}^{-1}$. In BDI and CII class 
\begin{align} \label{eq:TransformedQ}
& {\cal T} \begin{pmatrix} 
0 & Q({\bf k}) \\
Q^{\dagger}({\bf k}) & 0 
\end{pmatrix}
 {\cal T}^{-1}  \nonumber \\
 & = i \eta_0 U {\cal K}  
 \begin{pmatrix} 
0 & Q({\bf k}) \\
Q^{\dagger}({\bf k}) & 0 
\end{pmatrix}
 i \alpha \eta_0 U {\cal K} \nonumber  \\ 
 & = 
\begin{pmatrix} 
0 &  \alpha U Q^{*}({\bf k}) U^{*} \\
 \alpha U Q^{T}({\bf k}) U^{*} & 0
\end{pmatrix}, 
\end{align}
where we have used  ${\cal T}^2 = \alpha$ and ${\cal T}^{-1} = \alpha {\cal T}$. 
From Eq. \ref{eq:TransformedQ},  we read the transformed $Q'({\bf k})$ and ${Q'}^{\dagger}({\bf k})$ 
to be
\begin{equation} 
\begin{aligned}
& Q({\bf k}) \to Q'({\bf k})  = \alpha U Q^{*}({\bf k}) U^{*} \\ 
& Q^{\dagger}({\bf k}) \to {Q'}^{\dagger}({\bf k})  = \alpha U Q^{T}({\bf k}) U^{*}.\\ 
\end{aligned}
\end{equation}
From the above equation, the transformation of  $Q'^{-1}({\bf k})$  can be obtained 
\begin{align}
Q'^{-1}({\bf k}) = \alpha^{-1} U^{T} \left(Q^{-1}({\bf k})\right)^{*} U^{\dagger}. 
\end{align}
Now we can express the integral in Eq. \ref{eq:Wpinvariant} in terms of $Q({\bf k})$ by inserting 
the above expressions of  $Q'({\bf k})$ and  ${Q'}^{-1}({\bf k})$. We simplify the integrand 
as  follows 
\begin{equation}
\begin{aligned} \label{eq:IntegSimplification}
& \mathrm{Tr}\left({Q'}^{-1} \nabla_{\bf k} Q' \right) \\
& = \mathrm{Tr}\left(\alpha^{-1} U^{T}   \left(Q^{-1}({\bf k}))\right)^{*} U^{\dagger}  \nabla_{\bf k}
 \alpha U Q^{*}({\bf k}) U^{*} \right) \\
& = \alpha^{-1} \alpha ~ \mathrm{Tr}\left(U^{T} \left(Q^{-1}({\bf k}))\right)^{*}  \nabla_{\bf k} Q^{*}({\bf k}) U^{*} \right) \\
& =  \mathrm{Tr}\left(\left(Q^{-1}({\bf k})\right)^{*} \nabla_{\bf k} Q^{*}({\bf k}) U^{*} U^T \right) \\
& =  \mathrm{Tr}\left(\left(Q^{-1}({\bf k})\right)^{*}  \nabla_{\bf k} Q^{*}({\bf k})\right) \\
& =  \mathrm{Tr}\left(Q^{-1}({\bf k}) \nabla_{\bf k} Q({\bf k})\right)^{*}.
\end{aligned}
\end{equation}

\vspace{0.2cm}

In the above manipulation, we have used $U^{*} U^T = \left(U^{\dagger}U\right)^T = 1$, cyclic 
property of trace, and assumed the unitary matrix $U$ to be momentum independent. The above 
simplification implies  the following  important relation
\begin{align}
\textrm{Im}\left[\mathrm{Tr}\left({Q'}^{-1} \nabla_{\bf k} Q' \right)\right] = - \textrm{Im}\left[\mathrm{Tr}\left({Q}^{-1}
 \nabla_{\bf k} Q \right)\right].  
\end{align}
Therefore, we find that 
\begin{subequations}
\begin{align}
W' = &  \frac{1}{2\pi } \int d{\bf k} \cdot \textrm{Im}\left[\mathrm{Tr}\left({Q'}^{\dagger} \nabla_{\bf k} Q' \right) \right] \\
     = &  -\frac{1}{2\pi } \int d{\bf k} \cdot \textrm{Im}\left[\mathrm{Tr}\left({Q}^{\dagger} \nabla_{\bf k} Q \right) \right] \\
     = & -W
\end{align}
\end{subequations}
i.e. the topological charge $W$ of a WN is opposite to the charge carried by its 
time-reversed partner in class BDI and CII. 
A similar relation between the topological charges of two WNs which are related by 
the time-reversal symmetry can be established in class DIII and CI. Going through 
the same exercise as above, we find that 
\begin{align}
\textrm{Im}\left[\mathrm{Tr}\left({Q'}^{-1} \nabla_{\bf k} Q' \right)\right] =  \textrm{Im}\left[\mathrm{Tr}\left({Q}^{-1} 
\nabla_{\bf k} Q \right)\right].
\end{align}
in class DIII and CI. Therefore the topological charge $W$ of a WN has to be 
equal to the charge carried by its time-reversed partner in  class DIII and CI. 

%%%%%%%%%%%%%%%%%%%%%%%%%%%%%%%%%%%%%%%%%%%

%\vspace{0.5cm}

\bibliography{main}

\end{document}